\documentclass[aps,prl,twocolumn,superscriptaddress,raggedbottom,showpacs,floatfix]{revtex4-1}
\usepackage{amsmath}
\usepackage[next]{inputenc}
\usepackage[dvips]{epsfig}
\usepackage{bbm,bm,bbold}
\usepackage{booktabs}
\usepackage{multirow}
\usepackage{hhline}
\usepackage{comment}
\usepackage{float}

\usepackage{amsmath,amsfonts,amssymb,amsthm}
\usepackage{color}
\usepackage{graphicx,latexsym}
\usepackage{epstopdf}
\usepackage{comment}

\usepackage[next]{inputenc}
\usepackage[dvips]{epsfig}
\usepackage[colorlinks=true,citecolor=blue,linkcolor=blue]{hyperref}
\usepackage{bbm,bm}
\usepackage{booktabs}
\usepackage{multirow}
\usepackage{hhline}
\usepackage{comment}
\usepackage{float}

\usepackage{lipsum}

\usepackage{nameref}
\usepackage{varioref}
\usepackage{hyperref}
\usepackage{cleveref}

\usepackage{natbib}

\begin{document}
	\renewcommand{\vec}{\mathbf}
	\renewcommand{\Re}{\mathop{\mathrm{Re}}\nolimits}
	\renewcommand{\Im}{\mathop{\mathrm{Im}}\nolimits}

\preprint{APS/123-QED}

\title{Tunneling and Fluctuating Electron-Hole Cooper Pairs in Double Bilayer Graphene}% Force line breaks with \\
%\thanks{A footnote to the article title}%

\author{Dmitry K. Efimkin}
 \email{dmitry.efimkin@monash.edu}
 \affiliation{The Center for Complex Quantum Systems, The University of Texas at Austin, Austin, Texas 78712-1192, USA}
 
 \affiliation{School of Physics and Astronomy and ARC Centre of Excellence in Future Low-Energy Electronics Technologies, Monash University, Victoria 3800, Australia }

 \author{G. William Burg}%
 \affiliation{Microelectronics Research Center, Department of Electrical and Computer Engineering,
	The University of Texas at Austin, Austin, TX 78758, USA}

\author{Emanuel Tutuc}%
\affiliation{Microelectronics Research Center, Department of Electrical and Computer Engineering, The University of Texas at Austin, Austin, TX 78758, USA}

	\author{Allan H. MacDonald}
\affiliation{The Center for Complex Quantum Systems, The University of Texas at Austin, Austin, Texas 78712-1192, USA}

%\date{\today}% It is always \today, today,
             %  but any date may be explicitly specified

\begin{abstract} A strong low-temperature enhancement of the tunneling conductance between graphene bilayers has been reported recently, and interpreted as a signature of equilibrium electron-hole pairing, first predicted in bilayers more than forty years ago but previously unobserved.  Here we provide a detailed theory of conductance enhanced by fluctuating electron-hole Cooper pairs, which are a precursor to equilibrium pairing, that accounts for specific details of the multi-band double graphene bilayer system which supports several different pairing channels. Above the equilibrium condensation temperature, pairs have finite temporal 
coherence and do not support dissipationless tunneling. Instead they strongly boost the tunneling conductivity via  a fluctuational internal Josephson effect. 
Our theory makes predictions for the dependence of the zero bias peak in the differential tunneling conductance on temperature and electron-hole density imbalance that capture important aspects of the experimental observations.  In our interpretation of the observations, cleaner samples with longer disorder 
scattering times would condense at temperatures $T_c$ up to $\sim 50 {\rm K}$, compared to the record $T_c \sim 1.5 $K achieved 
to date in experiment. 
%\begin{description}
%\item[Usage]
%Secondary publications and information retrieval purposes.
%\item[Structure]
%You may use the \texttt{description} environment to structure your abstract;
%use the optional argument of the \verb+\item+ command to give the category of each item. 
%\end{description}
\end{abstract}

%\keywords{Suggested keywords}%Use showkeys class option if keyword
                              %display desired
\maketitle

%\tableofcontents

\section{I. Introduction}
\label{SecI}
The possibility of Cooper pairing in a system with spatially separated electrons and holes in 
semiconductor quantum wells was first anticipated more than forty years ago~\cite{LozovikYudson1,Shevchenko}.  
According to theory, strong Coulomb interactions allow pairing at elevated temperatures, 
which would provide a physical realization of dipolar superfluidity that is 
potentially relevant for applications.  The paired state is fragile however, and can be 
suppressed by disorder~\cite{EfimkinDisorder,MacDonaldDisorder}, or by Fermi-line mismatches due to 
the differences between electron and hole anisotropies~\cite{EfimkinLOFF,SeradjehLOFF} that are 
always present in conventional semiconductors~\cite{MacDonaldMismatch}. 
In fact equilibrium pairing has until recently been observed only in the presence of strong magnetic fields 
that quench the kinetic energies of electrons and holes and drive the system to 
the regime of strong correlations~\cite{EisensteinMacDonald, EisensteinReview}. 

\begin{figure}[t]
	\begin{center}
		\includegraphics[width=0.95\columnwidth]{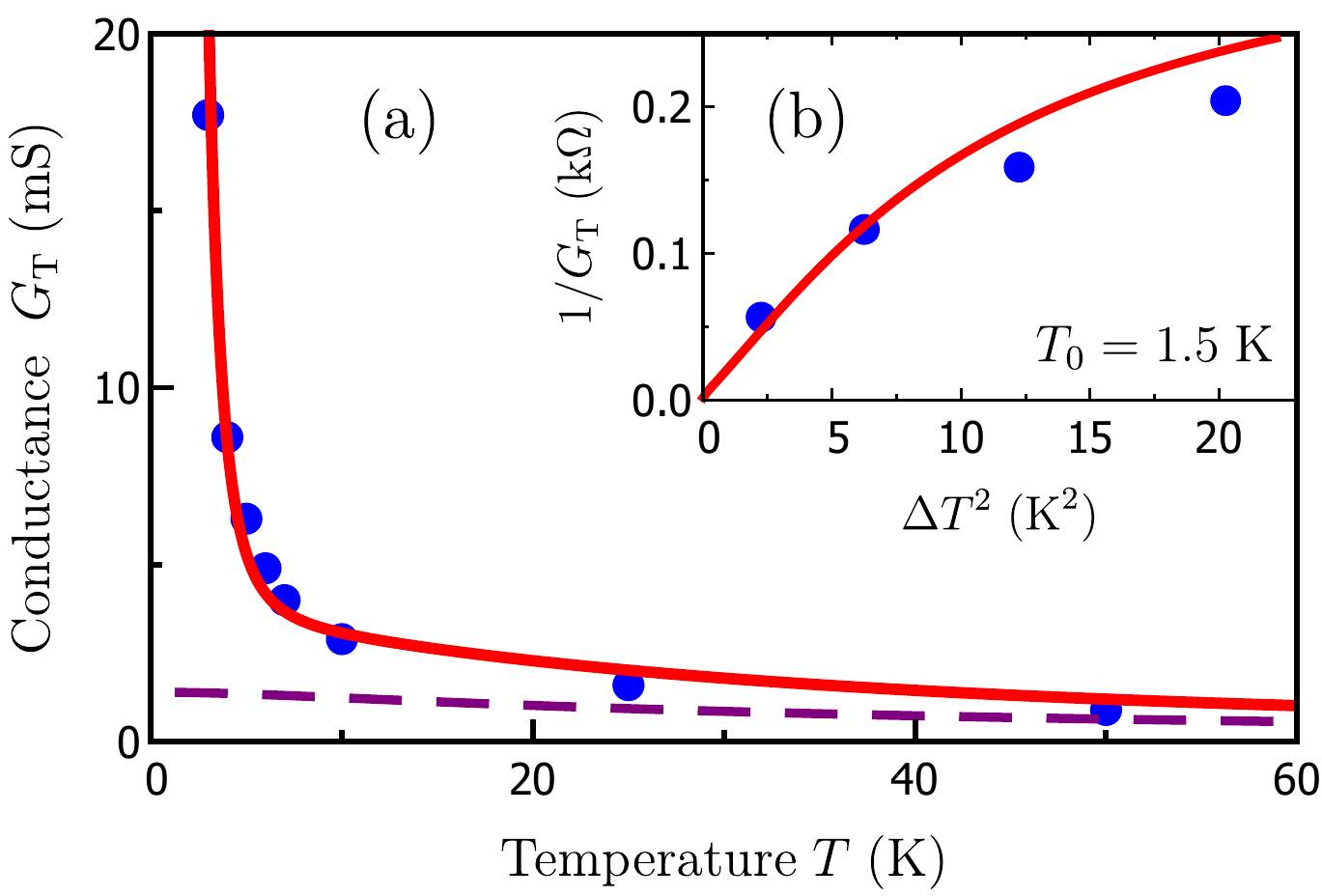}
		\caption{$\hbox{(a)}$ Temperature dependence of the tunneling conductance $G_\mathrm{T}$ between graphene bilayers at zero voltage bias ($V=0$). The red curve corresponds to the calculation that incorporates the effect of fluctuating Cooper pairs (Eq.~(\ref{GTIntra})) and accurately fits the experimental data presented by blue dots (Fig.~3-b in Ref.~\cite{TutucExperiment}). The purple dashed curve corresponds to the model of non-interacting electrons and holes (Eq.~(\ref{GTfit0})). The fitting details and parameters are presented in Sec.~V. Fluctuating Cooper pairs above the critical temperature $T_0=1.5\; \hbox{K}$ strongly enhance $G_\mathrm{T}$ and are responsible for its critical behavior $G_\mathrm{T}\sim (T-T_0)^{-2}$. The latter is derived and discussed in Sec.~IV-E and reasonably matches with the experimental data, as it is clearly seen in the inset \hbox{(b)}.}
		\label{Fig1}
	\end{center}\vspace{-1cm}
\end{figure}

Recent progress in fabricating single-atomic-layer two-dimensional materials has renewed interest in 
electron-hole pairing~\cite{GrapheneFirstLozovikSokolik, GrapheneFirstMacDonald, GrapheneFirstJoglekar, EfimkinMultiband, MacDonaldMultiband1, MacDonaldMultiband2, MacDonaldMultiband3,MacDonaldMultiband4, GrapheAfter1, GrapheAfter2, GrapheAfter3,ScreeningLargeT1, GrapheAfter4, GrapheAfter5,GrapheAfter6}.  Graphene-based two-dimensional 
electron systems not only have high mobility and almost perfect electron-hole symmetry but they make it possible to 
fabricate 
closely-spaced, and therefore strongly interacting, independently gated and contacted double layer structures. 
Very recently low-temperature enhancement of the tunneling conductance between graphene bilayers 
has been observed at matched concentrations of electrons and holes~\cite{TutucExperiment}. 
A typical conductance trace is presented in Fig.~\ref{Fig1}, where we see a tunneling conductance 
that appears to diverge at $T_0 \approx 1.5\;\hbox{K}$,
signaling equilibrium pair condensation.
This observation provides the first 
clear experimental signature of equilibrium electron-hole pair condensation in the absence of magnetic field~\cite{CommentCoulombDrag, CommentExcitonPolariton}.

Enhanced tunneling conductance has been observed previously in semiconductors bilayers in the strong field 
quantum Hall regime,~\cite{JEExperiment} and has been interpreted as an internal Josephson effect~\cite{EisensteinMacDonald}.
The differential conductance, does not diverge however, and instead has a sharp peak at zero bias.  
The property that the conductance peak width is smaller than temperature, and smaller than the single-electron 
scattering rate ({\it i.e.} the Landau level width) nevertheless points to a collective origin of the peak. 
Bilayers in the quantum Hall regime are predicted to support \emph{dissipationless} Josephson-like tunneling currents
in the presence of long-range electron-hole coherence~\cite{JEPrediction1,JEPrediction2}. 
The development of a quantitative theory of enhanced tunneling in quantum Hall systems~\cite{JETheory1,JETheory2,JETheory3,JETheory4} that fully explains the 
peak width has been challenged by the importance of inhomogeneity and disorder, and by strong interactions in the 
presence of dispersionless Landau levels.
Phase fluctuations that are inevitably present due to the two-dimensional 
Berezinskii-Kosterlitz-Thouless nature of the phase
transition~\cite{Berezinskii1,Berezinskii2,KosterlitzThouless1,KosterlitzThouless2} also play a role.
The theory of enhanced tunneling is simpler at zero magnetic field, at least in the weak coupling 
regime where the electron-hole pairing energy is small compared to the Fermi energy, allowing 
experiments to be explained more fully as we demonstrate below.

The enhancement of the tunneling conductance in the double bilayer graphene system has been observed over a wide density range $4\cdot10^{10}$$\,\sim\,$$10^{12}\,\hbox{cm}^{-2}$, where electronic correlations vary from moderate to weak~\cite{EHMonteCarlo}. 
The Bardeen-Cooper-Schrieffer (BCS) theory of electron-hole pairing has greater validity at weaker pairing. True internal Josephson behavior in this case  
occurs only below a critical temperature $T_0$
and is preceded by enhancement of the tunneling conductance that diverges as $T_0$ is approached 
as illustrated in Fig.~\ref{Fig1}. This critical behavior has been predicted by one of the 
authors~\cite{EfimkinJosephson} and has been interpreted as a \emph{fluctuational} internal Josephson effect. 	
It originates from partly coherent fluctuating electron-hole Cooper pairs~\cite{EfimkinJosephson,EfimkinDrag,RistVarlamovMacDonaldFazioPolini} that are a precursor of equilibrium pairing and reminiscent of Aslamazov-Larkin and related effects in superconductors~\cite{Varlamov,LarkinVarlamov,SkocpolTinkham}. 
Above $T_0$ fluctuating Cooper pairs have a finite coherence time~\cite{GrapheAfter1,Pseudogap1,Pseudogap2,Pseudogap3}
and cannot support a dissipationless tunneling current. 
While the 
recent observations do qualitatively agree with earlier theory, the double bilayer graphene system has some important differences compared to the double parabolic electron gas models considered previously.  These are related to the system's well known $2 \pi $ momentum space Berry phases.  We show here that accounting properly for these differences provides a better account of the low-temperature tunneling anomalies.   

In the present work we have developed a theory of the fluctuational internal Josephson effect in a system 
with closely spaced graphene bilayers. As we show below, the presence of valley and sublattice degrees of freedom 
provides three competing electron-hole channels for both intra-valley and inter-valley Cooper pairs. 
We show that three channels are nearly independent and have have different condensation temperatures and different sublattice structure.  The experimental enhancement of the tunneling conductance by fluctuating Cooper pairs  
can be explained only by the presence of competing channels that dominate in different 
temperature ranges.  In the vicinity of $T_0$, the tunneling conductance 
at zero bias is predicted to have a critical divergence $G_\mathrm{T}\sim (T-T_0)^{-2}$,
that matches the experimental data well as we see in Fig.~1. The calculated dependence of the tunneling conductivity on 	inter-layer voltage bias and carrier-density imbalance also match the experimental data~\cite{TutucExperiment} reasonably. We conclude that the observed enhancement of the tunneling conductance in double bilayer graphene is well explained by our fluctuational internal Josephson effect theory.  

The paper is organized as follows. In Sec. II we introduce a model that describes the low-energy physics of two closely spaced graphene bilayers. Sec. III is devoted to a description of fluctuating Cooper pairs above the critical temperature $T_0$. In Sec. IV we 
use these results as a starting point for a theory of the tunneling conductance. In Sec. V we compare our calculations
with the recent experimental data.  Finally in Sec. VI we discuss limitations of our theory and aspects of the experimental data that are still not well understood, and present our conclusions. 

\section{II. Model} \label{SecII}

\subsection{A. Weak and strong coupling regimes}

\begin{figure}[t]
	\begin{center}
		\vspace{-0.5cm}
		\includegraphics[trim=0.4cm 13.5cm 16cm 2.5cm, clip,width=1.0\columnwidth]{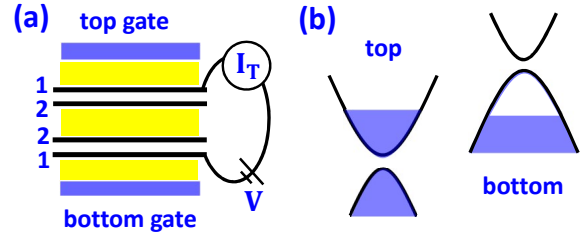}
		\vspace{-0.5cm}
		\caption{(a) Schematic of a double bilayer graphene system which specifies our sublayer labeling. External gates induce an excess of electrons (holes) in the top (bottom) bilayer and open a gap in the spectrum of isolated bilayers. (b) Electronic structure of the system in the case of the matched concentration of electrons and holes that most strongly favors electron-hole Cooper pairing. }
		\label{Fig2}
	\end{center}
\end{figure}

The system of interest contains two graphene bilayers separated by an insulator as sketched in Fig.~\ref{Fig2}-a. An external electric field perpendicular to the bilayers induces an excess of electrons in the top bilayer ($\mathrm{t}$) and their deficit in the bottom bilayer ($\mathrm{b}$). Inevitably, it also results in gaps $2|u|$ in isolated bilayer graphene spectra. The latter are in Fig.~\ref{Fig2}-b in the case of matched concentrations for electrons and holes, the most favorable regime for the Cooper pairing. Double bilayer physics is very rich and has been considered in a number of recent papers~\cite{GrapheAfter1,GrapheAfter2,GrapheAfter3,GrapheAfter4,EHMonteCarlo} that aim to provide realistic predictions of the critical temperature $T_0$ based on the microscopic model. Here we follow a different route	and consider a minimal phenomenological model that accounts for disorder and  describes the instability of the system towards electron-hole pairing in the weak coupling regime and can explain the observed enhancement of the tunneling conductance between graphene bilayers~\cite{TutucExperiment}.
	
The physics of electron-hole Copper pairing depends on three dimensionless parameter: Wigner-Seitz interaction strength parameter $r_\mathrm{s}=m e^2/\kappa \hbar^2 k_\mathrm{F}$ that scales the ratio of interactions and kinetic energy in an individual bilayer~\cite{CommentRs}, a parameter $k_\mathrm{F}d$ that scales the inter-layer Coulomb interactions respect to the intra-layer one, and a parameter $|u|/\epsilon_\mathrm{F}$ that determines the low-energy spectrum of individual bilayers. Here $k_\mathrm{F}$ and $\epsilon_\mathrm{F}$ are Fermi wave vector and energy for electrons and holes, $m$ is their effective constant, $\kappa$ and $d$ are a dielectric constant for the spacer between bilayers and its thickness. If $r_\mathrm{s}\ll 1,$ $k_\mathrm{F}d \gg 1$ and $|u|/\epsilon_\mathrm{F} \ll 1$ the system is in the weak coupling regime with pairing correlations only in the vicinity of Fermi level for electrons and holes. This regime can be described by the BCS theory for electron-hole pairing. The nature of the strong coupling regime ($r_\mathrm{s}\gg 1$ and $k_\mathrm{F} d \sim 1$) depends on the ratio $|u|/\epsilon_\mathrm{F}$. If $|u|\gg\epsilon_\mathrm{F}$ the state is a Bose-Einstein condensate (BEC) of indirect excitons that represent a bound state of electron and hole. It has been argued that the most favorable conditions for observation of electron-hole condensation are reached near the mid-point of the BEC-BCS crossover~\cite{GrapheAfter1,GrapheAfter2, GrapheAfter3, EHScreening, EHMonteCarlo}. In the opposite case the paired state is a multi-band BCS-like paired state~\cite{MacDonaldMultiband1,MacDonaldMultiband2,MacDonaldMultiband3,MacDonaldMultiband4,NeilsonNew} where pairing correlations also span to remote bands (valence band in the layer with excess of electrons and conduction band in the layer with excess of holes).

Enhanced conductance has been seen over a range of electron and hole densities that covers $4\cdot 10^{10} \sim 10^{12} \; \hbox{cm}^{-2}$ and corresponds to $r_\mathrm{s}=3.1\sim 0.72$ and $k_\mathrm{F} d=0.07\sim 0.35$~\cite{Estimations}. This range suggests to the transition from moderate to weak coupling regime occurs with increasing of electron and hole concentrations.  In the experimental setup~\cite{TutucExperiment} the Fermi energy $\epsilon_\mathrm{F}$ of charge carriers and the gap $2|u|$ in the spectrum of individual bilayers are not controlled independently, but with the same gate (that produces the electric field perpendicular to bilayers). While the density can be tuned within a wide range, the ratio $|u|/\epsilon_\mathrm{F}\approx d_\mathrm{BG}/d\approx 0.15$ is approximately fixed. Here $d_\mathrm{BG}$ is the thickness of individual bilayers. This favors the multi-band BCS-like state at strong coupling and extends the region of the applicability for the phenomenological weak coupling BCS theory that will be employed below.

\subsection{B. Phenomenological model in the weak coupling regime}

The spectrum sketched in Fig.~\ref{Fig2}-b has spin and valley degeneracy.  The spin degrees of freedom simply added a factor of $2$ in the tunneling conductance when the condensed state preserves spin-invariance and do not need to be treated explicitly. The low energy states in bilayer graphene are concentrated around two inequivalent valleys $K$ ($v=1$) and $K'$ ($v=-1$) situated at the corners of the first Brillouin zone, and are described by the two-band Hamiltonian~\cite{BilayerGrapheneReview}
\begin{equation}
H_0=\sum_\vec{p}\left[\hat{\psi}_{\mathrm{t} \vec{p}}^+ (\hat{h}_{\mathrm{t} \vec{p}} - \mu_\mathrm{t}) \hat{\psi}_{\mathrm{t}\vec{p}} + \hat{\psi}_{\mathrm{b} \vec{p}}^+ (\hat{h}_{\mathrm{b} \vec{p}} + \mu_\mathrm{b} ) \hat{\psi}_{\mathrm{b} \vec{p}}\right].
\end{equation}
where $\hat{\psi}_{\vec{p}}=\{\psi_{\mathrm{l}1 \vec{p}},\psi_{\mathrm{l} 2 \vec{p}}\}$ is a spinor of annihilation operators for electrons in both layers with sublattice indexes $\sigma=1,2$, that are numerated according to the sketch in Fig.~\ref{Fig2}-a. 
We assume that there is a deficit of electrons in the bottom layer, but it is instructive not to perform the transformation to field operators of holes.  
$\mu_\mathrm{t}=\epsilon_\mathrm{F}+h$ and $\mu_\mathrm{b}=\epsilon_\mathrm{F}-h$ characterize the electric potentials 
in the top and bottom bilayers. Here $\epsilon_\mathrm{F}$ is the average of the electron and 
hole Fermi energies at neutrality, while $2 h\ll \epsilon_\mathrm{F}$ is their  difference.
The matrices $\hat{h}_{\mathrm{t} \vec{p}}$ and $\hat{h}_{\mathrm{b} \vec{p}}$ are 
\begin{equation}
\hat{h}_{\mathrm{t} \vec{p}}=\begin{pmatrix}u & \frac{p_{\bar{v}_\mathrm{t}}^2}{2m} \\ \frac{p_{v_\mathrm{t}}^2}{2m} & -u \end{pmatrix}, \; \quad  \; \hat{h}_{\mathrm{b} \vec{p}}=\begin{pmatrix}-u & \frac{p_{v_\mathrm{b}}^2}{2m} \\ \frac{p_{\bar{v}_\mathrm{b}}^2}{2m} & u \end{pmatrix}. 	
\end{equation}
Here $v=\pm 1$ and $\bar{v}=\mp 1$ are valley indexes, $m$ is the electron mass, and $p_{v}=p_x+i v p_y$. The electric field perpendicular to bilayers opens a gap $2|u|$ separating conduction $\epsilon_{\mathrm{c}\vec{p}}=\epsilon_\vec{p}$ and valence $\epsilon_{\mathrm{v}\vec{p}}=-\epsilon_\vec{p}$  bands in each bilayer with $\epsilon_\vec{p}=\sqrt{u^2 +(p^2/2m)^2}$. In the weak coupling regime the pairing correlations appear in the vicinity of Fermi lines for electrons and holes, and the presence of remote bands (valence band in the layer with excess of electrons and conduction band in the layer with excess of holes) can be neglected. In this regime only the conduction band of the top bilayer and the valence band of the bottom bilayer
are relevant, and the corresponding spinor wave functions are 
\begin{equation}
\label{spinors}
|\mathrm{t} \mathrm{c} \vec{p} \rangle=\begin{pmatrix}\mathrm{c}_\vec{p} \;  e^{-i v_\mathrm{t} \phi_\vec{p}} \\ \mathrm{s}_\vec{p} \; e^{i v_\mathrm{t} \phi_\vec{p}}\end{pmatrix}, \;\; |\mathrm{b} \mathrm{v} \vec{p} \rangle=\begin{pmatrix}\mathrm{c}_\vec{p} e^{i v_\mathrm{b} \phi_\vec{p}} \\ - \mathrm{s}_\vec{p}  e^{-i v_\mathrm{b} \phi_\vec{p}}\end{pmatrix}.
\end{equation}
Here $\phi_\vec{p}$ is the polar angle; $c_\vec{p}=\cos(\vartheta_\vec{p}/2)$ and $s_\vec{p}=\sin(\vartheta_\vec{p}/2)$ with $\cos (\vartheta_\vec{p})=u/\epsilon_\vec{p}$. The spinors have valley dependent chirality $\pm v_{\mathrm{t}(\mathrm{b})} \phi_\vec{p}$ that defines a sublattice structure of fluctuating electron-hole Cooper pairs as will be shown below. 
We introduce disorder with the help of phenomenological scattering rates $\gamma_{\mathrm{t}(\mathrm{b})}$.   It 
is important in this theory to observe that because the electron and hole components of the 
Cooper pair are spatially separated and have opposite charges, both short-range and long-range Coulomb disorder lead to
pair-breaking~\cite{EfimkinDisorder,MacDonaldDisorder}.

\begin{figure}[t]
	\begin{center}
		\vspace{-0.5cm}
		\includegraphics[trim=4cm 15cm 7cm 3.7cm, clip, width=1.0\columnwidth]{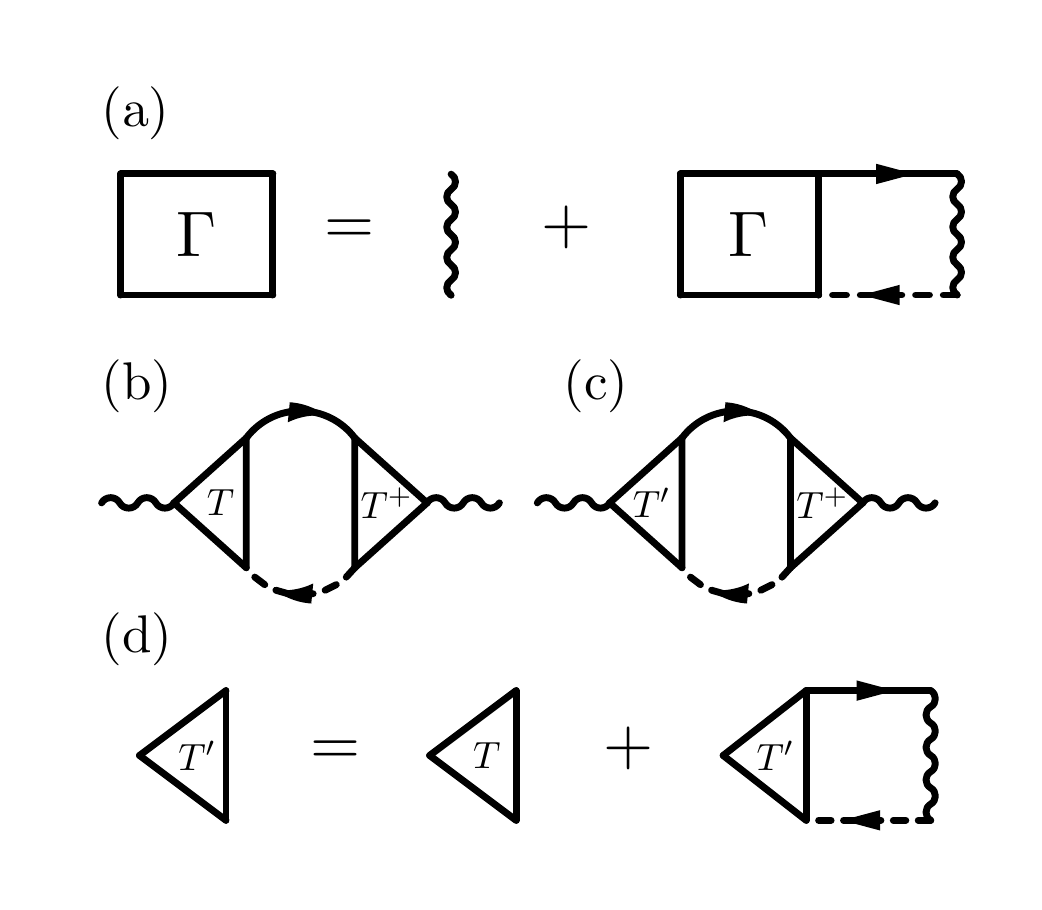}
		\vspace{-0.5cm}
		\caption{(a) Bethe-Saltpeter equation in the electron-hole channel. Divergence of the many-body vertex 
			$\Gamma(\omega=0,\vec{Q}_0)$ signals the double-bilayer electron-hole pairing instability with a momentum $\vec{Q}_0$. (b) The non-interacting  tunneling response function $\chi_0(\omega,\vec{q})$ introduced in Eq.~(\ref{GT1}). (c) The tunneling response function $\chi(\omega,\vec{q})$ with the renormalized vertex $t'$ that captures the effect of fluctuating Cooper pairs. (d) Renormalization of $t'$ that illustrates Eq.~(\ref{TRenorm}).}
		\label{Fig3}
	\end{center}
\end{figure}

In experiment~\cite{TutucExperiment} the relative angle $\theta$ between graphene bilayers can be adjusted. 
Since valleys $K$ and $K'$ reside at the corners of the first Brillouin zone, the valley momenta in the two layers do not 
match in the presence of a twist.
For momentum conserving tunneling, the current is maximized when the layers are aligned ($\theta=0$) or twisted by 
$\theta= n\pi/3$.  For even $n$, valley K (K') in one layer is aligned with valley K (K') in the other layer whereas 
for $n$ odd valley K (K') in one layer is aligned with valley K' (K) in the other layer.
When states are labeled by their momenta relative to the Brillouin-zone corners, the tunneling Hamiltonian
for $\theta$ close to $n\pi/3$ 
\begin{equation}
H_\mathrm{t}= T^++T=\sum_\vec{p}\left[\hat{\psi}_{\mathrm{b} \vec{p}}^+ \,\hat{t}^+\, \hat{\psi}_{\mathrm{t}\vec{p}+\vec{Q}} + \hat{\psi}_{\mathrm{t} \vec{p}+\vec{Q}}^+ \hat{t}\, \hat{\psi}_{\mathrm{b} \vec{p}}\right].
\end{equation}
Here $\vec{Q}=\vec{Q}_{\theta}+\vec{Q}_{\vec{B}}$ is the momentum splitting between valleys in the different layers.
The twist contribution at small relative angle $\theta\ll1 $ can be approximated as $\vec{Q}_{\theta}=-[\vec{q}_\mathrm{K}\times \vec{e}_\mathrm{z}] \theta$ where $\vec{q}_\mathrm{K}$ is the momentum for the Brillouin-zone corner in bilayer graphene. It has opposite directions for valleys $K$ and $K'$, and its magnitude is $q_\mathrm{K}=4\pi\hbar /3 a_0$ with $a_0$ the corresponding Bravais lattice period. The contribution induced by an in-plane magnetic field $\vec{B}_{||}$ is the same for two valleys and is equal to $Q_{\vec{B}}=ed [\vec{B}_{||}\times \vec{e}_\mathrm{z}]/\hbar c$. Because each bilayer is represented by a two-band model, the matrix $\hat{t}$ has four matrix elements which we treat in a phenomenological way below,
with the expectation that since $t_{22}$ 
corresponds to the tunneling between adjacent sublayers while $t_{11}$ to the tunneling between remote sublayers,
$|t_{11}|\ll |t_{12}|\approx |t_{21}|\ll |t_{22}|$.

\label{SecIII}
\section{III.Fluctuating Cooper pairs}
\subsection{A. The Cooper instability}
Due to Coulomb interactions between electrons and holes, the double-bilayer 
system is unstable towards Cooper pairing. 
Here we omit repulsive interactions within each graphene bilayer 
since its main effect in the considered weak coupling regime is a simple renormalization of the quasiparticle spectra.  The inter-bilayer attraction 
\begin{equation}
\label{HamiltonianInteractions}
H_\mathrm{int}=\sum_{\vec{p}\vec{p}'\vec{q}} \sum_{\sigma_\mathrm{t} \sigma_\mathrm{b}} U_\vec{q}\psi^+_{\mathrm{t},\vec{p}+\vec{q},\sigma_\mathrm{t}} \psi^+_{\mathrm{b}, \vec{p}'-\vec{q},\sigma_\mathrm{b}} \psi_{\mathrm{b} \vec{p}' \sigma_\mathrm{b}} \psi_{\mathrm{t} \vec{p} \sigma_\mathrm{t}}.
\end{equation}
Here $U_\vec{q}$ is the screened Coulomb potential estimated in our previous work~\cite{EfimkinMultiband,MacDonaldMultiband1,MacDonaldMultiband2}. 
Below we employ a multipole decomposition of the interaction and set the momenta magnitudes to the Fermi momentum so
that $U_\vec{q}$ reduces to a constant $U_l$ for each orbital angular momentum channel $l$. 
The set of $U_l$ parameters are also treated as phenomenological parameters 
with the expectation that the $s$-wave moment $U_\mathrm{s}\equiv U_0$ is largest. 
Since valleys are well separated in momentum space we neglect inter-valley scattering for electrons and holes.

The instability of double layer system towards electron-hole Cooper pairing is signaled by
divergence of the electron-hole channel scattering vertex  $\Gamma^{\sigma'_\mathrm{t} \sigma_\mathrm{t}}_{\sigma'_\mathrm{b} \sigma_{\mathrm{b}}}(\omega,\vec{p}',\vec{p},\vec{q})$ at zero frequency $\omega$. Note that in our approximation 
scattering conserves valley indices for electrons ($v_\mathrm{t}$) and holes ($v_\mathrm{b}$). 
The $\Gamma$-vertex satisfies the Bethe-Saltpeter equation presented in Fig.~\ref{Fig3}-a, and is
algebraic within the multipole approximation.  It is instructive to combine $\sigma_\mathrm{t}$ and $\sigma_\mathrm{b}$ 
into a single index $|\sigma_\mathrm{t} \sigma_\mathrm{b}\rangle$ that varies 
between $1$ and $4$ as $|1\rangle=|1 1\rangle$, $|2\rangle=|1 2\rangle$, $|3\rangle=|21\rangle$ and $|4\rangle=|2 2\rangle$. 
With this definition the Bethe-Saltpeter equation can be written (See Appendix A for details) in a compact matrix form:
\begin{equation}
\label{GammaEquation}
\hat{\Gamma}_{l'l}=U_{l} \, \delta_{l'l}+\sum_{l''} U_{l'} \,\hat{M}_{l'-l''} \,\Pi\, \hat{\Gamma}_{l'' l}. 
\end{equation}
Here momentum and frequency dependence are suppressed,
$l$ ($l'$) is the orbital momentum for the relative motion of two particles before (after) scattering, and 
$\hat{\Gamma}_{l'l}$ is the corresponding scattering matrix.  We have separated a factor of  $\Pi(\omega,\vec{q})$ which also appears as the single-step pair propagator in the Cooper ladder sum of a bilayer system without sub lattice degrees of freedom:
\begin{widetext}
	\begin{equation}
	\label{Pi}
	\Pi(\omega,\vec{q})=N_\mathrm{F}\left\{\ln\left[\frac{\epsilon_\mathrm{c}}{2\pi T}\right]-\frac{1}{2}  \sum_{\zeta=\pm} \left\langle \Psi\left(\frac{1}{2}+\frac{i[\omega + \zeta (h +  v_\mathrm{F} q \cos\phi_\vec{p} )] + \gamma}{4\pi T}\right)\right\rangle_{\phi_\vec{p}}\right\}.
	\end{equation}
\end{widetext}
Here $\epsilon_\mathrm{c}$ is an energy cutoff that is required for momentum-independent interactions. 
The average $\langle \cdot \cdot \cdot \rangle_{\phi_\vec{p}}$ is calculated respect to a polar angle $\phi_\vec{p}$. $\Psi(x)$ is the logarithmic derivative of $\Gamma$-function (or the digamma function), while $\zeta=\pm 1$ is the summation index.   $\gamma=\gamma_\mathrm{t}+\gamma_\mathrm{b}$ is the pair breaking rate,
which is the sum of the scattering rates for electrons $\gamma_\mathrm{t}$ and holes $\gamma_\mathrm{b}$. 
The expression for $\Pi(\omega,\vec{q})$ in Eq.~(\ref{Pi}) is 
well known~\cite{Varlamov,LarkinVarlamov} from previous work on systems without layer 
degrees of freedom.  The chiral nature of the bilayer graphene charge carriers 
is captured by the nontrivial matrix form-factor $\hat{M}_l$ in the Bethe-Salpheter Eq.~(\ref{GammaEquation}). 
The matrix form-factor $\hat{M}_l$ is defined as the multipole moment of the two-particle   
matrix element
\begin{widetext}
	\begin{equation}
	\label{FormFactorDef}
	\hat{M}^{\sigma'_t,\sigma_t}_{\sigma'_b,\sigma_b}(\phi_\vec{p})=\langle \sigma'_\mathrm{t}|\mathrm{t}\mathrm{c}\vec{p}+\frac{\vec{q}}{2}\rangle \langle \mathrm{t}\mathrm{c}\vec{p}+\frac{\vec{q}}{2}|\sigma_\mathrm{t}\rangle \langle \sigma_\mathrm{b}|\mathrm{b}\mathrm{v}\vec{p}-\frac{\vec{q}}{2}\rangle \langle \mathrm{b}\mathrm{v}\vec{p}-\frac{\vec{q}}{2}|\sigma'_\mathrm{b}\rangle.  
	\end{equation}
\end{widetext}
Corrections to the form-factor $\hat{M}_l$ due to finite Cooper pair momentum
$q$, $|\Delta M| = q^2  |u|/4 p_\mathrm{F}^2 \epsilon_\mathrm{F}$, 
are negligible in the weak coupling regime since $q\ll p_\mathrm{F}$. As a result, the matrix (\ref{FormFactorDef}) can be approximated as follows 
\begin{widetext}
	\begin{equation}
	\label{FormFactor}
	\hat{M}(\phi_\vec{p})=
	\begin{pmatrix}
	c^4 & -c^3 s e^{ -2 i v_\mathrm{b} \phi_\vec{p}}  &  c^3 s e^{-2 i v_\mathrm{t} \phi_\vec{p}}  & -c^2 s^2 e^{-2 i (v_\mathrm{t}+v _\mathrm{b}) \phi_\vec{p}} \\ 
	-c^3 s e^{2 i v_\mathrm{b} \phi_\vec{p}}  &  c^2 s^2 & -c^2 s^2 e^{2 i (v_\mathrm{b}-v_\mathrm{t}) \phi_\vec{p}} & c s^3 e^{-2 i v_\mathrm{t} \phi_\vec{p}}  \\ c^3 s e^{2 i v_\mathrm{t} \phi_\vec{p}} & -c^2 s^2 e^{2 i (v _\mathrm{t}-v_\mathrm{b}) \phi_\vec{p}} & c^2 s^2 & - c s^3 e^{-2 i v_\mathrm{b}\phi_\vec{p}} \\
	-c^2 s^2 e^{2 i (v_\mathrm{t}+v_\mathrm{b}) \phi_\vec{p}} & c s^3  e^{2 i v_\mathrm{t} \phi_\vec{p}} & - c s^3  e^{2 i v _\mathrm{b} \phi_\vec{p}} & s^4		
	\end{pmatrix} .
	\end{equation}
\end{widetext}
Here the coefficient $c$ and $s$ correspond to the coherence factors $c_\vec{p}$ and $s_\vec{p}$ in (\ref{spinors}) 
evaluated at the average Fermi energy $\epsilon_\mathrm{F}$ for electrons and holes and are given by 
\begin{equation}
\label{cs}
c^2=\frac{1}{2}\left (1+\frac{u}{\epsilon_\mathrm{F}}\right), \quad\quad s^2=\frac{1}{2}\left (1-\frac{u}{\epsilon_\mathrm{F}}\right). 
\end{equation}

The matrix form-factor (\ref{FormFactor}) shapes the sublattice structure of fluctuating electron-hole Cooper 
pairs in double-bilayer graphene.   Note that it couples scattering channels $\hat{\Gamma}_{l'l}$ with different orbital momenta.  
Importantly, $\hat{M}_l$ has only even harmonics $l=0,\pm 2, \pm 4$ that forbid scattering between states
with even and odd orbital momenta. For isotropic Coulomb interactions,
the $s$-wave moment $U_\mathrm{s}$ is expected to be largest. 
It is instructive to start by neglecting all other moments.  
In that case only $\hat{\Gamma}_{00}$ is nonzero. The latter depends on the $s$-wave moment of the form-factor $\hat{M}_0$ 
that has a different form for intra-valley and inter-valley Cooper pairs. We discuss these two case separately below.

\subsection{B. Intra-valley Cooper pairs}
For intra-valley ($v_\mathrm{t}=1$ and $v_\mathrm{b}=1$) electron-hole Cooper pairs the s-wave moment of the form-factor $\hat{M}_0=\langle \hat{M}(\phi_\vec{p}) \rangle_{\phi_\vec{p}}$ is given by
\begin{equation}
\label{MKK}
\hat{M}_0=
\begin{pmatrix}
c^4 & 0 &  0  & 0 \\ 
0  &  c^2 s^2 & -c^2 s^2 & 0  \\
0 & -c^2 s^2  & c^2 s^2 & 0 \\
0 & 0 & 0 & s^4		
\end{pmatrix} .
\end{equation}
The scattering problem decouples into the three channels
identified in Ref.~[\onlinecite{GrapheAfter3}] and the corresponding scattering 
vertex is given by

\begin{equation}
\label{GammaKK}
\frac{\hat\Gamma_{00}}{U_\mathrm{s}}=
\begin{pmatrix}
\frac{1}{L_{11}} & 0 &  0  & 0 \\ 
0  &  \frac{1-c^2 s^2 U_\mathrm{s}\Pi}{L_{12\hbox{-}21}} & \frac{-c^2 s^2 U_\mathrm{s} \Pi}{L_{12\hbox{-}21}} & 0  \\
0 & \frac{ - c^2 s^2 U_\mathrm{s}\Pi }{L_{12\hbox{-}21}}  &\frac{1-c^2 s^2 U_\mathrm{s}\Pi}{L_{12\hbox{-}21}} & 0 \\
0 & 0 & 0 & \frac{1}{L_{22}} 	
\end{pmatrix}.
\end{equation}
Here $L_\alpha=1 -\lambda^\mathrm{s}_\alpha \Pi/N_\mathrm{F}$ is a dimensionless inverse Cooper propagator for channel $\alpha$ and $\lambda^\mathrm{s}_\alpha$ is the corresponding coupling constant specified in Tab.~(\ref{TableIntra}).
$L_\alpha$ vanishes at the critical temperature $T_\alpha$ for 
the electron-hole pairing instability in channel $\alpha$.

\vspace{1mm}
\begin{table}[b]
	\caption{Coupling constants $\lambda_\alpha$ for intra-valley Cooper pairs}
	\label{TableIntra} 
	\begin{tabular}{|c|c|c|}
		\hline
		\quad Channel, $\alpha$ \quad  &\quad $s$-wave, $\lambda^\mathrm{s}$ \quad & \quad $d$-wave, $\lambda^\mathrm{d}$ \quad\\
		\hline
		11 & $c^4 N_\mathrm{F} U_\mathrm{s}$ & $2c^2 c^2 N_\mathrm{F} U_\mathrm{d}$ \\
		22 & $s^4 N_\mathrm{F} U_\mathrm{s}$ &  $2c^2 c^2 N_\mathrm{F} U_\mathrm{d}$ \\
		12\hbox{-}21 & $2s^2 c^2 N_\mathrm{F} U_\mathrm{s}$ & $(c^4 + s^4) N_\mathrm{F} U_\mathrm{d}$\\
		\hline
	\end{tabular}
	\label{TableIntravalley}
\end{table}

\begin{figure}[t]
	\begin{center}
		\includegraphics[width=0.9\columnwidth]{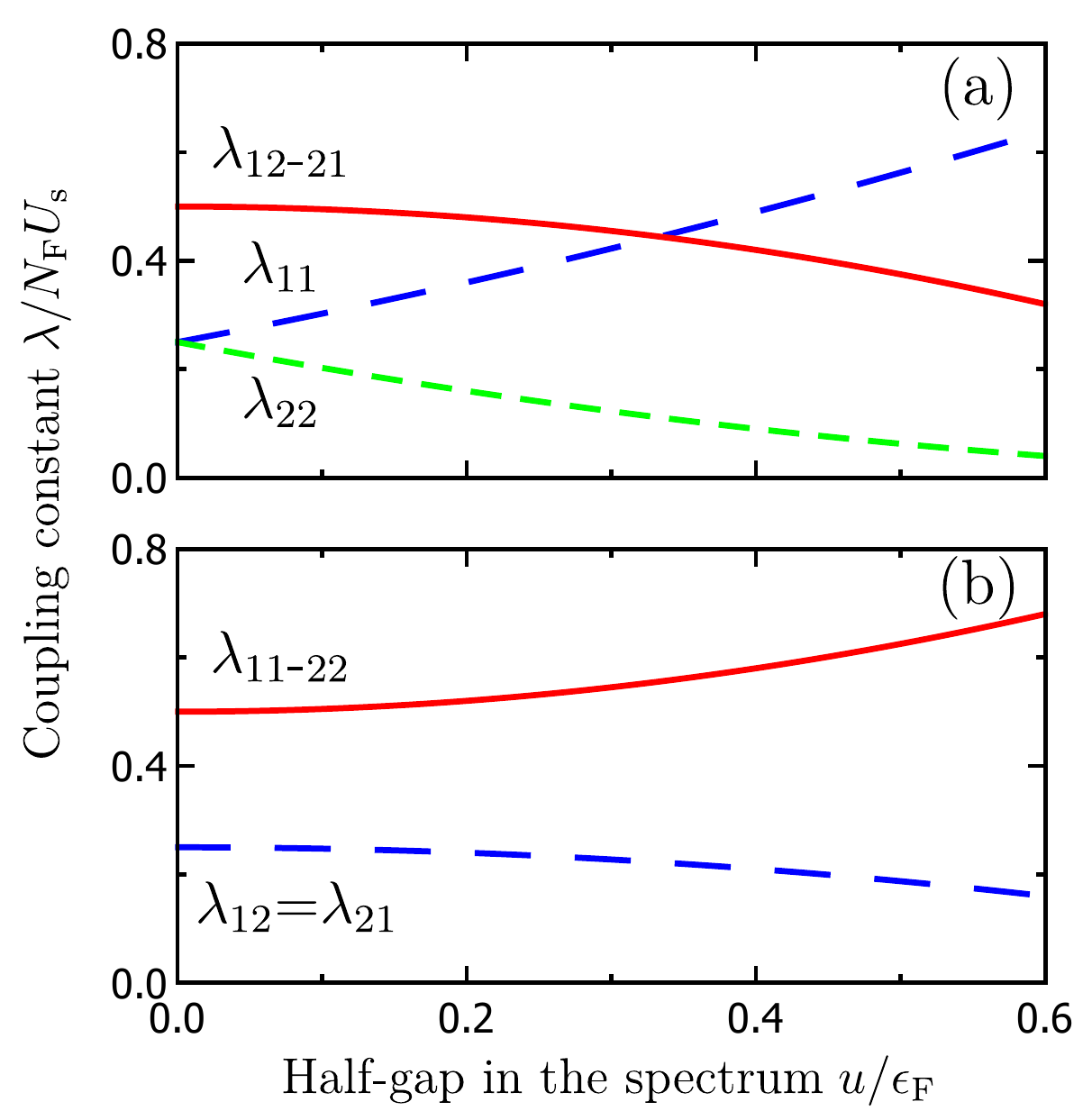}
		\caption{Dependence of coupling constants $\lambda_\alpha$ for different pairing channels on the displacement field parameter
				$u$.  For intra-valley ($\hbox{a}$) Cooper pairs a competition between channels is possible, but in the considered regime $|u|\ll\epsilon_\mathrm{F}$ the mixed one $12\hbox{-}21$ dominates. For inter-valley Cooper pairs ($\hbox{b}$) the hierarchy between channels does not depend on ratio between $|u|$ and $\epsilon_\mathrm{F}$ and the mixed channel $11\hbox{-}22$ is the dominant one.}
		\label{Fig3p5} 
	\end{center}
\end{figure}

In the absence of disorder and electron-hole density imbalances,
the critical temperatures are given by 
$\bar{T}_\alpha=2 e^\mathrm{C}\Lambda \exp[-1/\lambda^\mathrm{s}_\alpha]/\pi$, where $C=0.577$ is the Euler constant.
Although the coupling constant values $\lambda_\alpha$ can be fine-tuned by the displacement field, as it is presented in Fig.~\ref{Fig3p5}-a, their hierarchy is universal for the case 
$|u|\ll \epsilon_\mathrm{F}$. The coupling constant $\lambda_{12\hbox{-}21}^\mathrm{s}\approx 1/2$ is almost twice 
as large as the constants $\lambda_{11 (22)}^\mathrm{s}\approx 1/4$,
ensuring domination of the mixed channel $12\hbox{-}21$. 
Physically the presence of two sub lattice combinations ($|12\rangle$ and $|21\rangle$)
doubles the number of states that take part in the Cooper pairing.

\subsection{C. Inter-valley Cooper pairs} 
For inter-valley ($v_\mathrm{t}=1$ and $v_\mathrm{b}=-1$) electron-hole Cooper pairs the s-wave moment of the form-factor $\hat{M}_0=\langle \hat{M}(\phi_\vec{p}) \rangle_{\phi_\vec{p}}$ is given by
\begin{equation}
\label{MKKp}
\hat{M}_0=
\begin{pmatrix}
c^4 & 0 &  0  & -c^2 s^2 \\ 
0  &  c^2 s^2 & 0  & 0  \\
0 & 0  & c^2 s^2 & 0 \\
-c^2 s^2 & 0 & 0 & s^4		
\end{pmatrix} .
\end{equation}
This case also decouples into three independent channels, with scattering vertex 

\begin{equation}
\label{GammaKKp}
\frac{\hat{\Gamma}_{00}}{U_\mathrm{s}}=
\begin{pmatrix}
\frac{1-s^4 U_\mathrm{s} \Pi}{L_{11\hbox{-}22}} & 0 &  0  & \frac{- c^2 s^2 U_\mathrm{s} \Pi  }{L_{11\hbox{-}22}} \\ 
0  &  \frac{1}{L_{12}} & 0  & 0  \\
0 & 0  & \frac{1}{L_{21}} & 0 \\
\frac{-c^2 s^2 U_\mathrm{s} \Pi}{L_{11\hbox{-}22}} & 0 & 0 & \frac{1-c^4 U_\mathrm{s} \Pi }{L_{11\hbox{-}22}}		
\end{pmatrix}.
\end{equation}
Interestingly, the sublattice structure of the Cooper pairs and the corresponding coupling 
constants are different in intra-valley and inter-valley cases. 
The latter are presented in the Tab. (\ref{TableIntervalley}) and their dependence on displacement field parameter $u$ is shown in Fig.~\ref{Fig3p5}-b. In that case the hierarchy between coupling constants is universal and does not depend on the ratio between $|u|$ and $\epsilon_\mathrm{F}$. The mixed channel $11\hbox{-}22$ has the highest critical temperature.

\begin{figure}[t]
	\begin{center}
		\includegraphics[trim=0.1cm 3cm 0.1cm 3cm,clip,width=1.0\columnwidth]{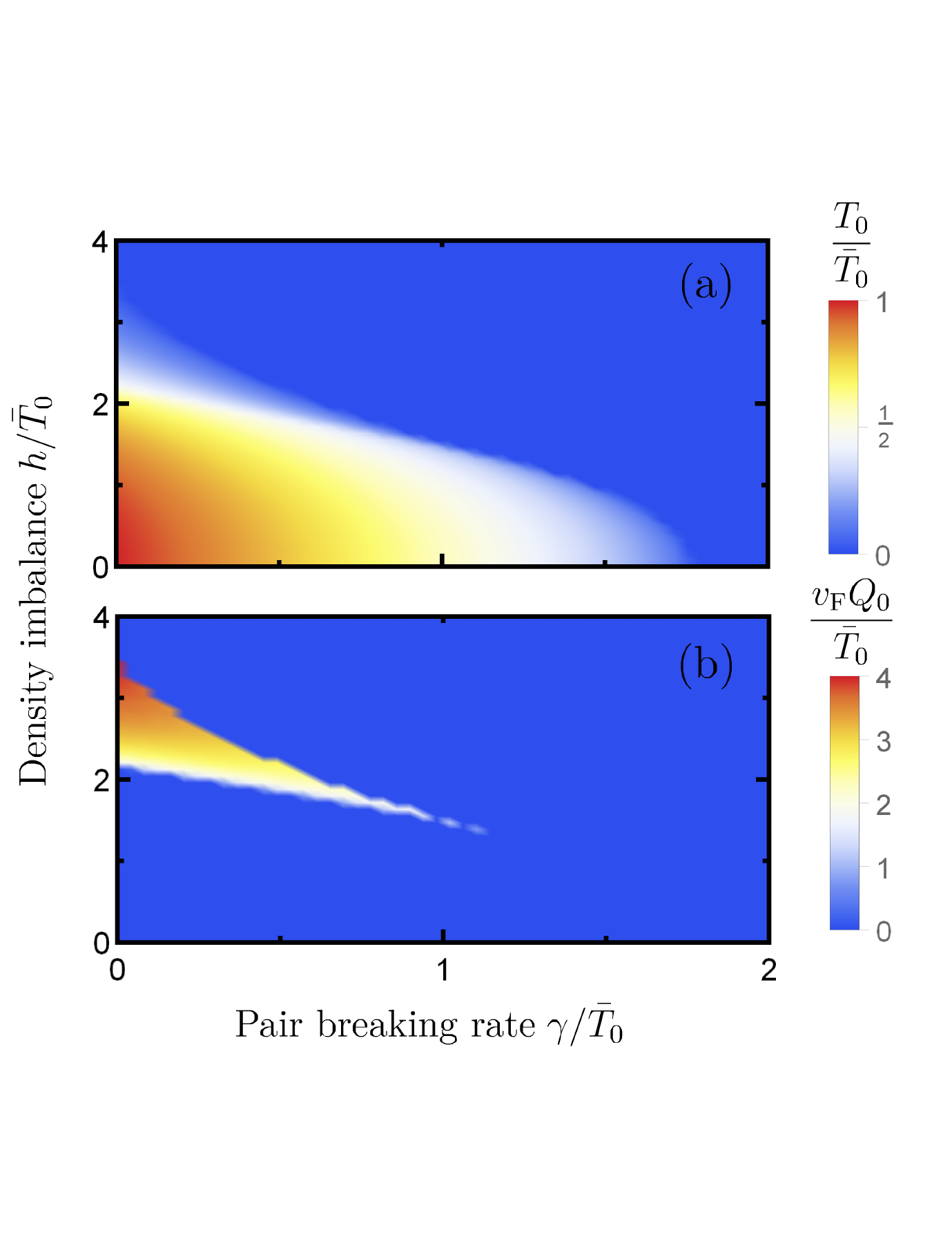}
		\caption{Dependence of the critical temperature $T_0$ (a) and the instability momentum of Cooper pairs $Q_0$ on the 
			pair-breaking rate $\gamma$ and electron-hole imbalance, parameterized by the difference between electron and hole Fermi energies $2h$.  When disorder is weak $\gamma\lesssim T_0$ the imbalance can stabilize the Fulde-Ferrell-Larkin-Ovchinnikov (FFLO) state with the finite Cooper pair momentum.}
		\label{Fig4}
	\end{center}
\end{figure}

\begin{table}[b]
	\caption{Coupling constants $\lambda_\alpha$ for inter-valley  Cooper pairs} 
	\begin{tabular}{|c|c|c|}
		\hline
		Channel, $\alpha$  & $s$-wave, $\lambda^\mathrm{s}$ & $d$-wave, $\lambda^\mathrm{d}$ \\
		\hline
		$12$ & $c^2 s^2 N_\mathrm{F} U_\mathrm{s}$ & $(c^4+s^4) N_\mathrm{F} U_\mathrm{d}$ \\
		$21$ & $c^2 s^2 N_\mathrm{F} U_\mathrm{s}$ & $(c^4+s^4) N_\mathrm{F} U_\mathrm{d}$ \\
		$11$-$22$ & $	(c^4+s^4) N_\mathrm{F} U_\mathrm{s}$ & $2 c^2 s^2 N_\mathrm{F} U_\mathrm{d}$\\
		\hline
	\end{tabular}
	\label{TableIntervalley}
\end{table}

The separation of scattering problem into three channels is not an artifact of the $s$-wave truncation,
but is maintained when higher multipole momenta of interactions $U_\mathrm{l}$ are taken into account. 
When the $d$-wave interaction $U_\mathrm{d}\equiv U_{\pm 2}$ is nonzero 
($s$- and $p$-wave momenta are decoupled and the latter is irrelevant) 
the scattering matrix $\Gamma_{l' l}$ is nonzero for $l=-2,0,2$. 
We show below that the only effect of the $d$-wave momentum $U_\mathrm{d}$ on the tunneling conductance between 
graphene bilayers is a renormalization of coupling constants 
$\lambda_\alpha=\lambda_\alpha^\mathrm{s}+\lambda_\alpha^\mathrm{d}$. 
The  $d$-wave coupling constants $\lambda_\alpha^\mathrm{d}$ are summarized in 
Tabs.~\ref{TableIntravalley} and ~\ref{TableIntervalley}.

\subsection{D. Disorder and density imbalances} 

Disorder and electron-hole density imbalances both reduce the critical temperature $\bar{T}_0$ (index $0$ corresponds to the channel $\alpha$ that has the largest critical temperature of electron-hole pairing).
The latter can stabilize the Fulde-Ferrell-Larkin-Ovchinnikov (FFLO) pairing state~\cite{FF,LO,SeradjehLOFF,EfimkinLOFF,Imbalance1,Imbalance2} with finite Cooper pair momentum $\vec{Q}_0$ (Ref.~\cite{Imbalance1,Imbalance2} also outlines a stabilization of the Sarma-phases~\cite{SarmaPhase} within the BEC-BCS crossover. These phases are unstable in the weak-coupling regime considered here).   The critical temperature $T_0$ and the 
instability momentum $\vec{Q}_0$ for a channel with highest critical temperature satisfy the  
equation $L_0(0,\vec{Q}_0)=0$, which can be recast as follows:
\begin{widetext}
	\begin{equation}
	\label{CPcrit}
	\ln\left[\frac{T_0}{\bar{T}_0}\right]+\frac{1}{2}  \sum_{\zeta=\pm} \left\langle\Psi\left(\frac{1}{2}+\frac{i\zeta ( h + v_\mathrm{F} Q_0 \cos\phi_\vec{p}) + \gamma}{4\pi T}\right)- \Psi\left(\frac{1}{2}\right) \right\rangle_{\phi_\vec{p}}=0.
	\end{equation}
\end{widetext}

The dependence of the critical temperature $T_0$ on the pair-breaking rate $\gamma$ and the
electron-hole imbalance expressed as a difference in Fermi energies, $h$, are illustrated in Fig.~\ref{Fig4}-a. 
The electron-hole pair instability is suppressed when the pair-breaking rate exceeds a 
critical value $\gamma\approx1.78\, \bar{T}_0$. If the rate does not exceed
$\gamma\approx T_0$ the Fulde-Ferrell-Larkin-Ovchinnikov (FFLO) state with finite Cooper pair momentum $\vec{Q}_0$
is stabilized by finite imbalance. 
The dependence of the instability momentum $\vec{Q}_0$ is presented in Fig.\ref{Fig4}-b. Its magnitude can be approximated by $v_\mathrm{F} Q_0\approx h$, corresponding to the difference between the Fermi momenta for electrons and holes.

\subsection{E. Coherence time and length of fluctuating Cooper pairs} 

In the absence of electron-hole imbalance,
the Cooper propagator $L_0^{-1}$ of the dominating channel at small frequencies and momenta simplifies to
\begin{equation}
\label{CooperL}
L_0^{-1}(\omega,\vec{q})=\frac{1}{\lambda_0}\frac{1}{i \omega \tau-\epsilon_{\vec{q}}}, \quad \quad  \epsilon_{\vec{q}}= \epsilon + \frac{\xi^2 \vec{q}^2}{\hbar^2}.
\end{equation}
Here $\epsilon=\ln\left[T/T_0\right]$ is the energy scale that is required to create a uniform fluctuating Cooper pair. It vanishes at the critical temperature $T_0$ and is linear $\epsilon\approx\Delta T/T_0$ in its vicinity $\Delta T\approx T_0$. Here $\Delta T=T-T_0$. $\tau$ and $\xi$ are characteristic time and spatial scales for Cooper pairs that are given by 
\begin{equation}
\label{CoherenceTimeLenght}
\tau=\frac{\hbar \Psi'\left(\frac{1}{2}+\frac{\gamma}{4\pi T}\right)}{4 \pi T}, \, \, \, \xi=\frac{\hbar v_F |\Psi''\left(\frac{1}{2}+\frac{\gamma}{4\pi T}\right)|^{\frac{1}{2}}}{8 \pi T}.
\end{equation}
They are connected to the coherence time and length of Cooper pairs by $\tau^*=\tau/2 \epsilon$ and $\xi^*=\xi/\sqrt{\epsilon}$ that 
diverge at the critical temperature for electron-hole condensation $T_0$. 
The Cooper propagator (\ref{CooperL}) has its only pole on the imaginary frequency axis at
$\omega_{\vec{q}}=-i /2\tau^*_\vec{q}$ with $\tau_\vec{q}^*=\tau/2\epsilon_\vec{q}$,
reflecting the dissipative nature of Cooper pairs dynamics. 
Due to the presence of a finite temporal coherence time $\tau^*$,
fluctuating Cooper pairs do not provide a dissipationless Josephson current, 
but do strongly enhance the tunneling conductance at zero voltage bias.

\section{IV. Tunneling conductivity}
\label{SecIV}
\subsection{A. Linear response theory for the tunneling conductance} 
When inter-layer tunneling amplitudes are treated in leading order of perturbation theory, 
the inter bilayer tunneling conductance at finite voltage bias $V$ is~\cite{Mahan,Tunneling2D1,Tunneling2D2}
\begin{equation}
\label{GT1}
G_\mathrm{T}(V)=\frac{8 A e^2}{\hbar}  \frac{\Im[\chi(eV,\vec{Q})]}{eV}.
\end{equation}
Eq.~(\ref{GT1}) accounts for the fourfold degeneracy due to the presence of valley and spin degrees of freedom 
and $A$ is the sample area. Here $\chi(\Omega)$ is the retarded correlation function 
corresponding to the imaginary-time ordered correlation function constructed from tunneling operators
\begin{equation}
\chi(\tau,\vec{q})=-\langle T_\mathrm{M}T(\tau,\vec{q})T^+(0,\vec{q})\rangle.
\end{equation}
$\chi(\omega,\vec{q})$ can be constructed from $\chi(\tau,\vec{q})$ by the 
usual Fourier transform and analytical continuation steps. 
Without Coulomb interactions between electrons and holes the response function $\chi(\omega,\vec{q})$ corresponds to the single electron-hole loop diagram depicted in Fig.~\ref{Fig3}-b and is given by 
\begin{equation}
\label{Chi0}
\chi_0(\omega,\vec{q})=\hat{t}^+ \hat{M}_0 \, \Pi \, \hat{t}, 
\end{equation}
Here $\hat{t}=\{t_{11},t_{12},t_{21},t_{22}\}$ is the vector of tunneling matrix elements in the compact representation,
and $\Pi\equiv \Pi(\omega,\vec{q})$ is a single step pair propagator in the 
Cooper ladder defined in Eq.~(\ref{Pi}). 
In the presence of Coulomb interactions, the single-particle Green functions and the tunneling vertex need to be renormalized. 
The renormalization of Green functions results in a dip of the density of states at the 
Fermi level~\cite{RistVarlamovMacDonaldFazioPolini}, that does not produce any singularities in the tunneling conductivity 
and thus is unimportant and can be neglected. 
The renormalized tunneling vertex $\hat{t}\rightarrow \hat{t}'\equiv\hat{t}'(\omega,\vec{q})$ diverges at the critical temperature 
$T_0$ and is responsible for the drastic enhancement of the tunneling conductivity in its vicinity. 
The renormalization of tunneling vertex is presented in Fig.~\ref{Fig3}-d and the corresponding 
equation for $\hat{t}'$ can be written as
\begin{equation}
\label{TRenorm}
\hat{t}'_l=t \, \delta_{l0}+\sum_{l'} U_l \, \hat{M}_{l-l'}\, \Pi\;\hat{t}_{l'}'. 
\end{equation}
The matrix form-factor $\hat{M}$ couples even orbital channels and we neglect all multipole moments 
except for $s$- and $d$.   (The $p$-wave multipole moment is decoupled from the $s$-wave one and is irrelevant.)
Since the form-factors $\hat{M}$ are different for intra-valley and inter-valley Cooper pairs, we again consider these two 
cases separately.

\subsection{B. Intra-valley tunneling}  

Without Coulomb interactions between electrons and holes the response function $\chi(\omega,\vec{q})$ is a sum of three non-interfering terms that correspond to three channels $\alpha$ introduced in Sec.~III
and identified in Ref.~[\onlinecite{GrapheAfter3}]:
\begin{equation}
\label{Chi0KK}
\chi^0=(c^4 |t_{11}|^2+ s^4|t_{22}|^2 + |t_{12} -t_{21}|^2 c^2 s^2)\, \Pi
\end{equation}
The three channels are not coupled by Coulomb interactions and are renormalized independently as follows:
\begin{equation}
\label{ChiKK}
\chi=\left(\frac{c^4 |t_{11}|^2}{L_{11}} + \frac{s^4|t_{22}|^2}{L_{22}} +  \frac{|t_{12} -t_{21}|^2 c^2 s^2}{L_{12\hbox{-}21}}\right)\Pi.
\end{equation}
Here $L_\alpha$ is the inverse Cooper propagator for each channel and $\lambda_\alpha=\lambda_\alpha^\mathrm{s}+\lambda_\alpha^\mathrm{d}$ is the corresponding coupling constant. Importantly the only role of the $d$-wave interaction moment is the renormalization the coupling constant $\lambda_\alpha$.
\subsection{C. Inter-valley tunneling} 

When opposite valleys are aligned by a twist angle between bilayers close to 
$\theta=\pi/3$, the response function $\chi(\omega,\vec{q})$ for noninteracting electrons and holes is 
\begin{equation}
\label{Chi0KKp}
\chi_0=(c^2s^2 |t_{12}|^2+ c^2s^2|t_{21}|^2 + |c^2t_{11} - s^2 t_{22}|^2)\Pi.
\end{equation}
It again is a sum of three non-interfering terms that correspond to three channels $\alpha$. Coulomb interactions do not couple 
the channels but renormalize as follows 
\begin{equation}
\label{ChiKKp}
\chi=\left(\frac{c^2s^2 |t_{12}|^2}{L_{12}}+ \frac{c^2s^2|t_{21}|^2}{L_{21}} + \frac{|c^2t_{11} - s^2 t_{22}|^2}{L_{11\hbox{-}22}}\right)\Pi.
\end{equation}
Each channels acquires its own Cooper propagator $L_\alpha$ with the coupling 
constant $\lambda_\alpha=\lambda_\alpha^\mathrm{s}+\lambda_\alpha^\mathrm{d}$. 

\subsection{D. Tunneling conductance}
Due to the remarkable conservation of momentum for electron tunneling, the conductance $G_\mathrm{T}$ is observable only if graphene bilayers are aligned with twist angle $\theta=0$ or $\theta=\pi/3$. In the former case the tunneling is intra-valley, while in the latter case it is inter-valley. Therefore we refer to the cases $\theta=0$ and $\theta=\pi/3$ as to intra-valley and inter-valley alignments. As we explain below, there important differences in the enhancement of tunneling conductance by fluctuating Cooper pair in these two cases. 

Within the model of non-interacting electrons and holes the  tunneling conductance is dominated by tunneling between adjacent sublayers $t_{22}$ and can be approximated (both for intra- and inter-valley alignments) as follows
\begin{equation}
\label{GTfit0}
G_\mathrm{T}^0=\frac{8 A e^2}{\hbar} \frac{ s^4 |t_{22}|^2\mathrm{Im}[\Pi(eV,\vec{Q})]}{eV}.
\end{equation}
In the presence of interactions, the enhancement of tunneling conductance by fluctuating Cooper pairs  works very differently for intra- and inter-valley alignments. The reason is a drastic difference in the sublattice structure for the pairing correlations.

For intra-valley fluctuating Cooper pairs, the dominating channel $12\hbox{-}21$ does not involve pairing correlations at adjacent sublayers. As result, the corresponding contribution of the channel $12\hbox{-}21$ is weaker than that of $22$, since the latter involves tunneling between adjacent sublayers, except in the vicinity of the critical temperature $T_0$. The tunneling conductance is governed by the competition of two channels and can be approximated as follows
	\begin{widetext}
		\begin{equation}
		\label{GTIntra}
		G_\mathrm{T}= \frac{8 A e^2}{\hbar} \left(  \frac{s^4|t_{22}|^2}{|L_{22}|^2} +  \frac{|t_{12} -t_{21}|^2 c^2 s^2}{|L_{12\hbox{-}21}|^2} \right) \frac{\mathrm{Im}[\Pi(eV,\vec{Q})]}{eV}.
		\end{equation} 
	\end{widetext}
As we explain in the next section, the competition between channels is essential for the explanation of the experimental data~\cite{TutucExperiment}, that in our interpretation is strongly depend on mutual orientation of graphene bilayers.

For the inter-valley alignment, the dominant channel for fluctuating Cooper pairs involves pairing correlations at adjacent sublayers. As a result, the tunneling conductance can be well approximated by a single term corresponding to the channel $11\hbox{-}22$ and is given by    
	\begin{widetext}
		\begin{equation}
		\label{GTInter}
		G_\mathrm{T}= \frac{8 A e^2}{\hbar}  \frac{s^4|t_{22}|^2}{|L_{11\hbox{-}22}|^2} \frac{\mathrm{Im}[\Pi(eV,\vec{Q})]}{eV}.
		\end{equation} 
	\end{widetext}
The inter-valley alignment case has not been studied experimentally yet, but according to our theory is most favorable for observations of the fluctuational internal Josephson effect.

\subsection{E. Critical behavior of tunneling conductance}
At matched concentrations of electrons and holes the voltage dependence of tunneling conductance $G_\mathrm{T}$ in the vicinity of the critical temperature $T_0$ acquires a Lorentzian shape which is governed by the factor 
\begin{equation}
\label{Critical}
F(eV,\vec{Q})= \frac{\mathrm{Im}[\Pi]}{eV |L_0|^2}=\frac{\tau}{(eV\tau)^2 + \epsilon_\vec{Q}^2}.
\end{equation}
Here $\epsilon_\vec{Q}=\epsilon+\xi^2 \vec{Q}^2/\hbar^2$ and $\epsilon=\ln (T/T_0)$ can be interpreted as an energy of fluctuating Cooper pairs. In the absence of a valley splitting $\vec{Q}=0$, the amplitude of the peak has a long high temperature tail $F(0,0)=\tau/\ln^2[T/T_0]$. It diverges near the critical temperature $T_0$ as a function of $\Delta T=T-T_0$ in the critical manner as $F(0,0) \approx \tau T_0^2/\Delta T^2$ with the index $2$. Its width at half maximum $e V_\mathrm{HM}=1/ \tau^*=2 \Delta T /T_0 \tau$ is equal to the inverse coherence time $\tau^*$ of fluctuating Cooper pairs and vanishes linearly at $T_0$.  In the presence of a valley splitting that can be induced by in-plane magnetic field or relative twist, fluctuating Cooper pairs with finite momentum  $\vec{Q}$ are probed in tunneling experiments. Temperature dependence of the peak width is modified as $eV_{HM}=\left(1+(\xi^* \vec{Q})^2 \right)/\tau^*$. It has a universal form as a function of coherence time $\tau_*=2\tau/\epsilon$ and length $\xi^*=\xi/\sqrt{\epsilon}$ for fluctuating Cooper pairs that allow to extract them form the experimental data in the presence of in-plane magnetic field.

For the case of the inter-valley alignment ($\theta=\pi/3$), the temperature dependence of the tunneling conductance~(\ref{GTInter}) is well approximated by Eq.~(\ref{Critical}) in a wide temperature range. For the case of intra-valley alignment ($\theta=0$) channels $12\hbox{-}21$ and $22$ compete with each other. As a result, tunneling conductance~Eq.~(\ref{GTIntra}) can not be approximated by a simple analytical expression over as wide temperature range. It is well approximated by the critical behavior given by Eq.~(\ref{Critical}) only in the narrow temperature range where the contribution of the channel $21\hbox{-}21$ dominates.

\section{V. Comparison with experiment}
\label{SecV}
In Ref.~[\onlinecite{TutucExperiment}] the twist angle between graphene layers can in principle be tuned
to access the intra-valley ($\theta=0$) and inter-valley ($\theta=\pi/3$) tunneling cases,
although experimental results for tunneling conductance $G_\mathrm{T}$ 
have so far been reported only for one alignment.  
A preliminary analysis of the data suggests that a divergent zero bias peak appears on the top of a background 
with a weaker temperature dependence. This behavior can be explained by the competition between channels $12\hbox{-}21$ and $22$. This is not surprising since top and bottom graphene bilayers originate from the same flake, and are aligned with the twist angle $\theta = 0$. Calculations for the intra-valley alignment are presented below and compared with the experimental data, while ones for the inter-valley one are presented in Appendix B. 

To fit the experimental data, the tunneling conductance $G_\mathrm{T}$ has been calculated with the help of Eq.~(\ref{GTIntra}). The model has a large number of fitting parameters. To adjust their values and compare our results with the experimental data we use the following strategy that involves six steps.

1) The experimental data has already been carefully analyzed~\cite{TutucExperiment} 
	within a model of noninteracting charge carriers. In the wide range of concentrations the non-interacting model explains the tunneling data very well except in the case of opposite polarity charge carriers with nearly equal electron and hole densities.  Based on the fits to experimental data away from matched electron and hole densities we can confidently assign values for the adjacent layer tunneling amplitude, $|t_{22}|=30\; \mu \hbox{eV}$, and the disorder-broadening 
	energies $\gamma_{\mathrm{t}(\mathrm{b})}=4 \; \hbox{meV}$.  Note that the measured disorder broadening parameter $\gamma_{\mathrm{t}(\mathrm{b})}$ is much larger than $T_{0}$, where $T_{0}$ is the temperature at which the tunneling conductance appears to diverge experimentally.  It is immediately clear therefore, even before performing a detailed analysis, that the condensation temperature must be strongly suppressed by disorder.

2) The colossal enhancement of the tunneling conductance has been observed in the wide density range $4\cdot 10^{10} \sim 10^{12} \hbox{cm}^{-2}$. The strength of the interactions in the system decreases  with doping and we have chosen the elevated doping level $n=7.4 \cdot 10^{11}\; \hbox{cm}^{-2}$ because of the detailed experimental data is available in this case. This doping level corresponds to the average concentration of electrons and holes $n=(n_\mathrm{e}+n_\mathrm{h})/2$, while the electron-hole imbalance $\Delta n \ll n$ can be present.

3) The transport experiments with similar double bilayer graphene samples and with similar gating geometry have already been performed~\cite{TutucExp1}. The doping level of charge carriers $\epsilon_\mathrm{F}$ and the gap $2|u|$ in the electronic spectrum of bilayer graphene are independent and are controlled by the same gate. At the considered average density $n=7.4 \cdot 10^{11}\; \hbox{cm}^{-2}$, the effective mass parameter of bilayer graphene can be approximated as $m\approx0.04\; m_0$, while the gap is equal to $2|u|\approx 6.6\;\hbox{meV}$ and is much smaller than the corresponding Fermi energy $\epsilon_\mathrm{F}\approx 20\;\hbox{meV}$. It follows that $|u|/\epsilon_\mathrm{F}\approx 0.16$, which implies sublayer polarization within the bilayers $c^2\approx 0.58$ and $s^2\approx 0.42$ to be modest.

4) The bare critical temperature without disorder $\bar{T}_0\approx 50 \; \hbox{K}$ can be recalculated from the actual critical temperature at the considered doping level $T_0=1.5 \; \hbox{K}$ and the Cooper pair-breaking rate $\gamma=\gamma_\mathrm{t}+\gamma_\mathrm{b}=8\; \hbox{meV}$ that has been chosen above with help of Eq.~(\ref{CPcrit}).  The system is in the regime of strong pair breaking $T_0 \ll \bar{T}_0 \sim \gamma$ 	and the value $\gamma/T_0\approx 1.74$ is very close to the critical value $1.78$. 	Experimentally, singular behavior of the tunneling conductance is observed only in the cleanest samples.

5) The bare critical temperature $T_0$ of Cooper pairing in the weak-coupling regime is given by $\bar{T}_0=2 e^\mathrm{C}\Lambda \exp[-1/\lambda_0]/\pi$, where $C=0.577$ is the Euler constant. Here $\lambda_0=\lambda_0^\mathrm{s}+\lambda_0^\mathrm{d}$ corresponds to the mixed-channel $12\hbox{-}21$ that dominates in the considered regime $|u|\ll \epsilon_\mathrm{F}$. Approximating the high energy cutoff as $\epsilon_\mathrm{c}\approx 2 \epsilon_\mathrm{F}$ and employing the values of $\bar{T}_0$ and $\epsilon_\mathrm{F}$ chosen above we get $\lambda_0\approx0.44$. The applicability condition for the weak-coupling approach $\lambda_0\ll 1$ is not well satisfied, but this value for $\lambda_0$ corresponds to moderate coupling regime $\lambda_0\lesssim 1$ that justifies the approximations used in our theory.

\begin{figure}[t]
	\begin{center}
		\includegraphics[width=1.0\columnwidth]{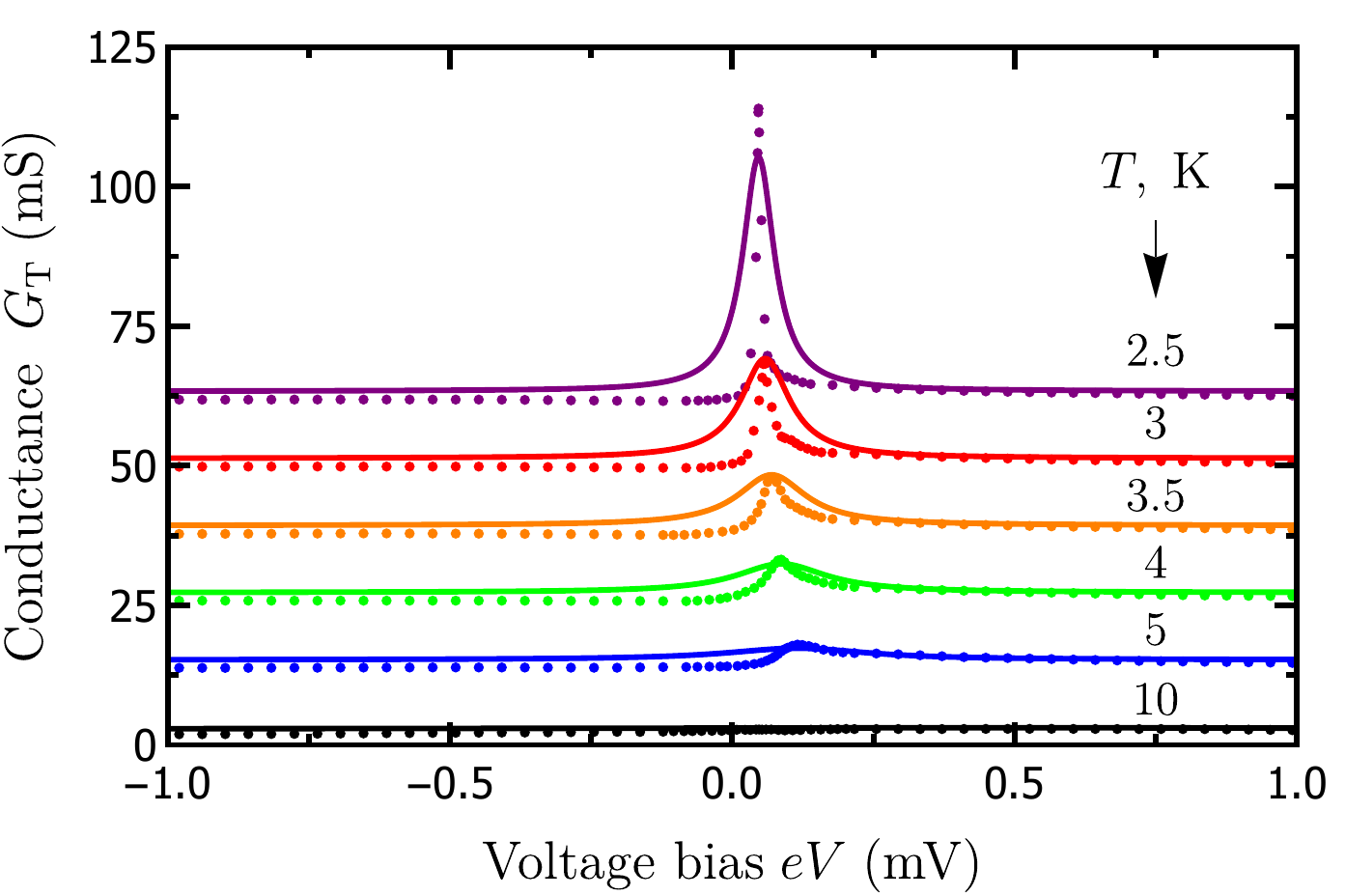}
		\caption{Dependence of the tunneling conductance $G_\mathrm{T}$ on bias voltage $V$ at different temperatures $T$. The concentrations of electrons and holes do match. The solid curves correspond to calculations, while the dotted lines are experimental data (Fig.~S2-b in Ref.~\cite{TutucExperiment}).  For clarity the curves at adjacent temperatures are offset by $\Delta G_\mathrm{T}=12 \; \hbox{mS}$. The theory that incorporates the effect of fluctuating Cooper pairs (Eq.~(\ref{GTIntra})) reasonably fits temperature dependence of the peak height, but overestimates its width  and does not capture its asymmetry. Possible origins of the asymmetry are discussed in Sec.~VI.}
		\label{Fig5} 
	\end{center}
\end{figure}

6) The relative contribution to the tunneling conductance $G_\mathrm{T}$ of channels $12\hbox{-}21$ and $22$ can be characterized by a dimensionless parameter $r=|t_{12}-t_{21}|^2/ |t_{22}|^2$ that we treat in the phenomenological way. Along with the sample area $A$, $r$ and $A$ are only free parameters of the model that have not yet been assigned. We adjust them by fitting the measured temperature dependence for tunneling conductance at zero voltage bias $V=0$ (and also at matched concentrations of electrons and holes and zero in-plane magnetic field), that is presented in Fig.~\ref{Fig1}, with the theory that incorporates the effect of fluctuating electron-hole Cooper pairs (Eq.~\ref{GTIntra}). The area $A$ can be be obtained by matching the high temperature behavior of $G_\mathrm{T}$ since the contribution of $12\hbox{-}21$ in this case is negligible small. It results in $A\approx 397 \;\mu \hbox{m}^2$. The value of $r\approx 7.6 \cdot  10^{-5}$ is obtained by fitting the singular behavior of the tunneling conductance in the vicinity of the critical temperature. The corresponding theoretical curve is also presented in 
	Fig.~\ref{Fig1} and matches with the experimental data reasonably well over a wide temperature range. 
	
In Fig.~\ref{Fig1} we also present calculations within the model of noninteracting electrons and holes (Eq.~(\ref{GTfit0})). This model severely underestimates the tunneling conductance $G_\mathrm{T}$ in the case of matched electron and hole concentrations, where interactions are crucial to explain enhanced tunneling conductance at low temperature and singular behavior in the vicinity of the critical temperature $T_0$ for Cooper pair condensation. We will keep all parameters chosen above in further calculations and investigate an impact of finite voltage bias between bilayers $V$, electron-hole imbalance $\Delta n$, and in-plane magnetic field $B$ at tunneling conductance. We will also present only results of the theory that incorporates the effect of fluctuating Cooper pairs.

\begin{figure}[t]
	\begin{center}
		\includegraphics[trim=0cm 3.3cm 0cm 1.2cm, clip, width=0.98\columnwidth]{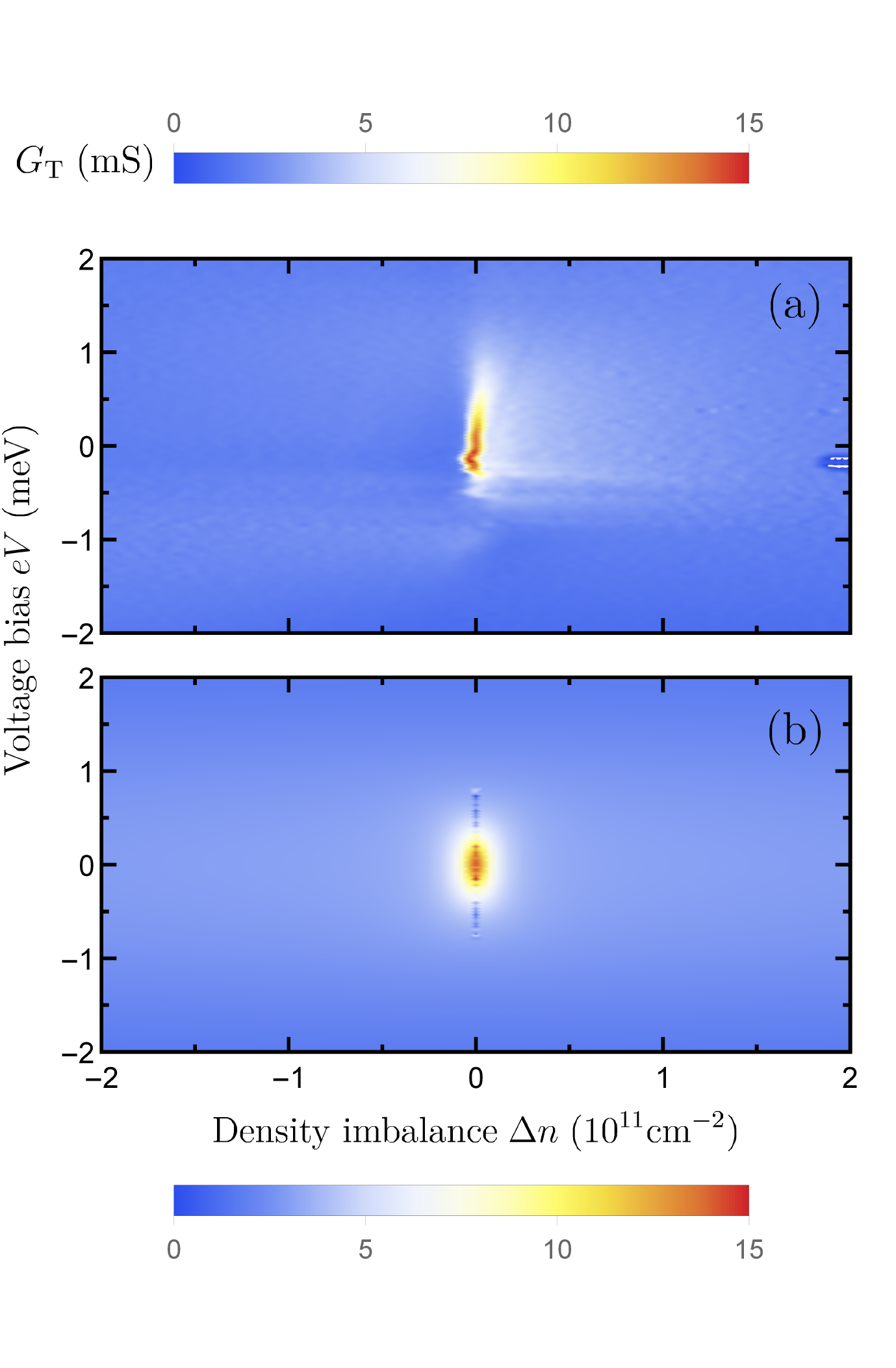}
		\caption{Dependence of the tunneling conductance $G_\mathrm{T}$ on voltage bias $V$ and electron-hole density imbalance $\Delta n$. Their average density $n=(n_\mathrm{e}+n_\mathrm{h})/2=7.4 \cdot 10^{11} \hbox{cm}^{-2}$ is fixed and the temperature is $3.5 \; \hbox{K}$. The top subplot (a) corresponds to experiment(Fig.~5-c in Ref.~\cite{TutucExperiment}), and the bottom one (b) to theory. The cut of this plot at zero voltage bias $V=0$ is presented as Fig.~\ref{Fig6}.}
		\label{Fig7}
	\end{center}
\end{figure}

A comparison between theory and experiment for the 
voltage dependence of the tunneling conductance between graphene bilayers
$G_\mathrm{T}$ is presented in Fig.~\ref{Fig5}. The zero bias peaks emerge 
with a decreasing temperature on a top of a smooth background that corresponds to the channel $22$. The width of the background $eV_\mathrm{HM}$ is governed by the single-particle energy scales $2 \pi T$ and $\gamma$ and is approximately equal to the largest of them. The width of the zero bias peak in the vicinity of $T_0$ is much smaller than the single-particle disorder scale, that demonstrates its collective origin. While the temperature dependence of peak height is well fit by the theory, the width dependence is captured only qualitatively and is overestimates by the factor of 2. The experimental data also exhibit a voltage asymmetry that becomes more prominent at low temperatures. 
Within our phenomenological model an asymmetrical voltage dependence of the tunneling conductance 
can be obtained if the scattering rates of electrons and holes $\gamma_{\mathrm{t}(\mathrm{b})}$ that define Cooper pair scattering rate as $\gamma=\gamma_\mathrm{t}+\gamma_\mathrm{b}$ are energy-dependent. 
The simple linear dependence $\gamma=\gamma+\gamma' \omega$ with a phenomenological parameter $\gamma'$ does not 
capture the observed asymmetry however.  A quantitative understanding of the asymmetry requires a microscopic 
understanding of disorder mechanisms that is outside the scope of the present work.

\begin{figure}[t]
	\begin{center}
		\includegraphics[width=1.0\columnwidth]{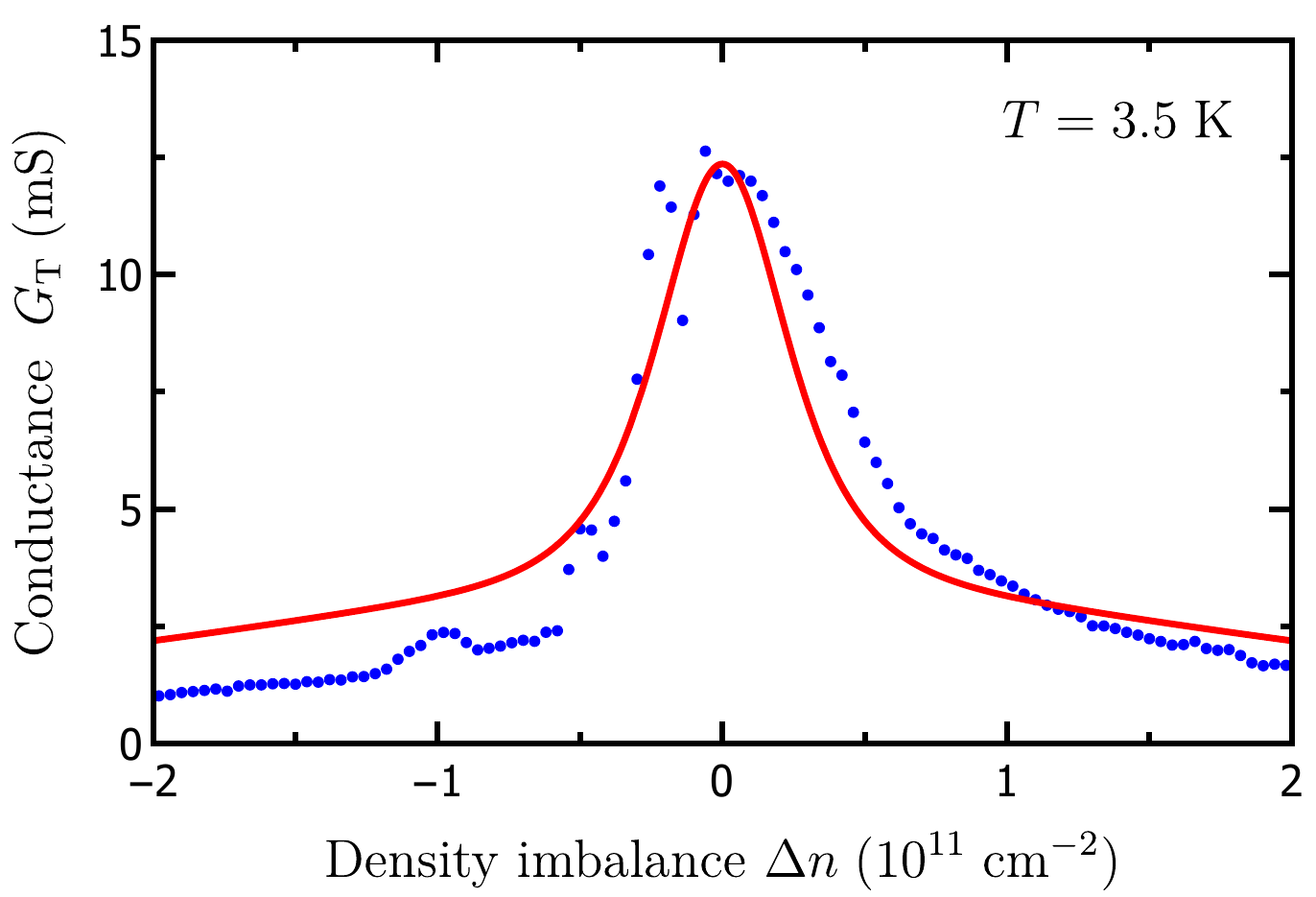}
		\vspace{-0.5cm}
		\caption{The dependence of tunneling conductance $G_\mathrm{T}$ at zero voltage bias $V=0$ on the electron-hole density imbalance $\Delta n$. Their average e density $n=(n_\mathrm{e}+n_\mathrm{h})/2=7.4 \cdot 10^{11} \hbox{cm}^{-2}$ is fixed and the temperature is $3.5 \; \hbox{K}$. The solid curve is theory and the dotted curve is experimental data extracted from the $V=0$ line from Fig.~5-c in Ref.~\cite{TutucExperiment}. The theory captures the curve profile reasonably well, but does not capture the asymmetry. Possible origins of the asymmetry are discussed in Sec.~VI.}
		\label{Fig6}
	\end{center}
\end{figure}

\begin{figure}[t]
	\begin{center}
		\includegraphics[trim=0.1cm 0cm 0.1cm 0.1cm, clip, width=0.98\columnwidth]{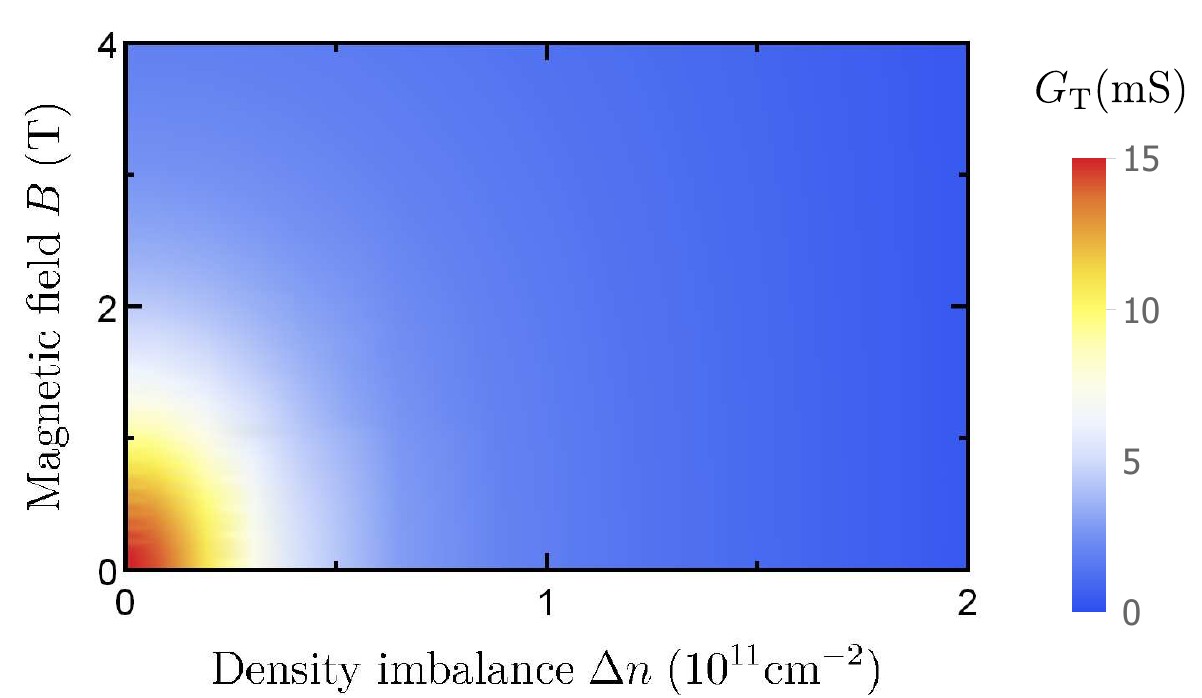}
		\caption{Theoretical dependence of tunneling conductance $G_\mathrm{T}$ on in-plane magnetic $B$ and electron-hole density imbalance $\Delta n$. Their average density $n=(n_\mathrm{e}+n_\mathrm{h})/2=7.4 \cdot 10^{11} \hbox{cm}^{-2}$ is fixed and the temperature is $3.5 \; \hbox{K}$. The dependence of $G_\mathrm{T}$ is smooth and featureless that demonstrates for the set of parameters used for fitting (the regime of strong pair breaking $\gamma \gtrsim T_0$) the FFLO state with finite Cooper pair momentum is not stabilized by the density imbalance $\Delta n$.}
		\label{Fig8}
	\end{center}
\end{figure}

The comparison between theory and experiment for the dependence of the tunneling conductance 
$G_\mathrm{T}$ on the electron-hole 
imbalance $\Delta n$ and the voltage bias $V$ are 
summarized in Fig.~\ref{Fig7} and a zero-bias-voltage cut of the comparison 
made in this color plot at zero voltage bias is presented in Fig.~\ref{Fig6}. 
The theoretical curves again agree reasonably well with the data. 
A density imbalance splits the Fermi lines of the electrons and holes and disfavors their Cooper pairing.
As  seen in Fig.~\ref{Fig4}, the FFLO state with a finite Cooper pair momentum can not be stabilized in the
strong pair breaking regime realized in the experiment. 
The critical temperature $T_0$ is maximal for Cooper pairs with zero Cooper pair momentum 
and decreases monotonically in the presence of imbalance and vanishes if the latter exceeds the critical temperature. 
As a result, the dependence of tunneling conductance on imbalance $\Delta n$ and in-plane magnetic filed $B$ that is 
presented in Fig.~\ref{Fig8} is smooth and featureless. 
Fig.~\ref{Fig8} presents theoretical curves since the experimental data 
for magnetic field and density-balance dependence at this temperature is not yet
available.  The theory qualitatively explains the decrease of peak height with magnetic field 
studied experimentally at lower temperatures $T_0\approx 1.5\;\hbox{K}$, 
but considerably overestimates its effect.  In this case the system is in the paired state whose behavior lies 
outside the range of validity of the 
present theory.  Other less fundamental limitations might also explain this discrepancy 
as we discuss in more details in the next section. We conclude, that our theory of the fluctuational internal Josephson effect, combined with specific features of the 
multiple-channel structure of pairing in double bilayer graphene provides a reasonable 
overall description of experiment.

\section{VI. Discussions}\label{sec:SecVI}
Our theory of the internal fluctuational Josephson effect does not account for interactions between fluctuating Cooper pairs. The Gaussian nature of the theory we employ is more clearly seen within an alternate derivation of the tunneling conductance. The latter employs the auxiliary field approach and is presented in Appendix C.  Interactions between fluctuating pairs can be safely omitted in the wide range of temperatures $\Delta T\sim T_0$, and are important only in the critical region $\Delta T_{Gi} \lesssim Gi \, T_0$ where fluctuations are large and strongly interfere with each other. 
Here $Gi=T_0/E_F$ is the Ginzburg number calculated in Appendix D.  The double bilayer graphene system studied
experimentally~\cite{TutucExperiment} is close to the weak coupling regime,
and the critical region $\Delta T_\mathrm{Gi}=T_0^2/E_\mathrm{F}\approx 10 \; \hbox{mK}$ is much smaller than the 
temperature range $\Delta T\approx 4 \; \hbox{K}$ where the tunneling conductance is strongly enhanced. 
The picture of noninteracting fluctuating Cooper pairs is therefore well justified to address the basic phenomenon 
identified in experiment.  

Fluctuating Cooper pairs in conventional superconductors alter the thermodynamics of the normal state 
only in the critical region $\Delta T_{Gi}.$  The conductivity and the magnetic susceptibility~\cite{Varlamov}
are however not only singular at the critical temperature, but 
have long high temperature tails~\cite{SkocpolTinkham}. The high temperature tail for the diamagnetic susceptibility $\chi \sim \chi_\mathrm{L}/\ln^2[T/T_0]$, where  
$\chi_\mathrm{L}$ is the Landau diamagnetic susceptibility in the normal state, was predicted~\cite{AslamazovlarkinDiamagnetsim,BulaevskiiDiamagnetsim,MakiDiamagnetsim} 
theoretically and observed in experiments~\cite{SkocpolTinkham}. 
The reason is  the paired state is superconducting and mediates the perfect diamagnetism 
that makes even a small number of fluctuating Cooper pairs important. 
Similarly the equilibrium paired state of spatially separated electrons and holes provides in fluctuational internal Josephson effect that 
colossally enhances the inter-layer tunneling.  That is why even a small density of fluctuating Cooper pairs can make a strong impact on the tunneling conductance above the critical temperature $T_0$ which also has the
high-temperature tail proportional to $1/\ln^2[T/T_0]$ that is clearly seen in Fig. 1.

In the vicinity of the critical temperature $T_0$ zero bias peak shape is universal and governed by the factor $F(eV,\vec{Q})$ given by Eq.~(\ref{Critical}). It can be rewritten as $F(eV,\vec{Q})  =\mathrm{Im}[L_0^{-1}(eV,\vec{Q})]/eV$. Thus an imaginary part of the Cooper propagator $L_0^{-1}(eV,\vec{Q})$, that can be interpreted as Cooper pair susceptibility~\cite{CommentSusceptibility}, is directly probed in tunneling experiments~\cite{CommentCooperPropagator}. It should be noted that Cooper pair-susceptibility of a superconductor can also be probed in tunneling Josephson junction in which one side is near its critical temperature while the other is well below its critical temperature~\cite{Ferrell,Scalapino}. The junction does not support dissipationless Josephson tunneling current, but the tunneling current at finite voltage is strongly enhanced by fluctuating Cooper pairs that grow in the vicinity of $T_0$. The latter has been observed experimentally~\cite{SCPairingexp1,SCPairingexp2}.

The enhancement of intra-valley tunneling that corresponds to the alignment ($\theta=0$) has only been reported so far. In this case the instability happens in one channel, while the main contribution to the tunneling conductance comes form the different one. As a result the divergent contribution to the conductance due to fluctuating Cooper pairs appears on the top of background with weak temperature dependence that dominates at higher temperatures. We predict an enhancement of inter-valley tunneling ($\theta=\pi/3$) to be much more profound because Cooper pairs in the dominating pairing channel involve electrons and holes settled at adjacent sublayers. We present calculations for this case in Appendix B.

The Fulde-Ferrell-Larkin-Ovchinnikov (FFLO) state with finite Cooper pair momentum has been predicted in conventional superconductors more than sixty years ago~\cite{FF,LO}. It requires the splitting of Fermi surfaces/lines for pairing electrons with opposite spins. The splitting can be induced by a magnetic field provided that its paramagnetic effect is larger than its diamagnetic one (Chandrasekhar-Clogston limit ~\cite{Chandrasekhar,Clogston}).  This condition is rarely satisfied even in layered conventional superconductors subjected to in-plane magnetic field. There are few observations in heavy-fermion and organic superconductors
where FFLO state signatures have been claimed but are still debated (See Ref.~\cite{FFLOReviewCndesedMatter} and Refs.~\cite{FFLOReviewColdAtoms1,FFLOReviewColdAtoms2} for reviews of progress in solid state and cold atoms systems). 
So far the FFLO state has not been unambiguously identified.  In double bilayer graphene the densities of electrons and holes can be controlled separately in a way that opens the FFLO state up for experimental study in a condensed matter system~\cite{EfimkinLOFF,SeradjehLOFF}.  The FFLO can be unambiguously identified if it appears 
from the dependence of the zero-bias peak on imbalance and in-plane magnetic field,
since the latter makes it possible to probe Cooper pairs with finite momentum. 
In the vicinity of the instability to the uniform paired state, the tunneling conductance monotonically decreases with in-plane 
magnetic field (as presented in Fig.~\ref{Fig8}).  In the vicinity of an instability to the FFLO  state 
the tunneling conductance achieves a maximum at finite field-induced momentum shift 
$\vec{Q}_\mathrm{B}=\vec{Q}_0$ where $\vec{Q}_0$ is the corresponding momentum of Cooper pairs. 
We discuss how distinguish these states in more details in Appendix B, where calculations for the inter-valley tunneling are presented. 

The sensitivity of Cooper pairing to a disorder opens a possibility of a granular electron-hole state in the presence of its strong long-range variations. In this state the pairing happens in disconnected or weakly coupled regions with minimal amount of disorder and does not support the spatial coherence. It makes the transport properties of the system including Coulomb drag effect to be different from ones in the uniform paired state. A tunneling conductance in the granular state is still colossally enhanced since the latter requires temporal coherence of Cooper pairs but not the spatial one. 

The interpretation of experiment provided by our theory suggested that the pairing critical 
temperature would be substantial if samples with weaker disorder could be fabricated.  This finding is perhaps 
a bit surprising since the experiments are for the most part conducted in the weak to moderate coupling regime ($r_\mathrm{s}=0.72\sim3.3$) where some researchers have argued that critical temperatures should be strongly suppressed by screening, especially accounting for spin and valley degeneracy~\cite{KharitonovEfetov, MacDonaldMultiband3} (See also arguments that this approach considerably underestimates the critical temperature~\cite{MacDonaldMultiband1,MacDonaldMultiband2,GrapheAfter1,GrapheAfter2}).  Our theory also suggests that high pairing temperatures should be achievable in double single-layer graphene systems, since there is nothing in 
its structure that puts single layers at a disadvantage relative to bilayer. Moreover, for the linear spectrum in monolayer graphene the Wigner-Seitz radius is defined in a different way  $r_\mathrm{s}=2.19/\kappa$ and could achieve even larger value $r_s=1.1$ if hBN is used as a spacer between graphene sheets. Here $\kappa$ is the corresponding dielectric constant. Future experimental work which seeks to weaken pair-breaking by disorder and which explores double single-layer graphene systems as well, is therefore important.  

In summary, the theory of the fluctuational internal Josephson effect developed here explains the anomalies 
in the tunneling conductance between graphene bilayers observed experimentally at equal electron and hole densities,
including their dependence on temperature, bias voltage bias and electron-hole imbalance. 
Some aspects of the observations are nevertheless not understood.  
First of all the observed dependence of the tunneling conductance on bias voltage has an asymmetry 
between positive and negative bias that becomes more prominent with decreasing temperature. 
At first glance the asymmetry is unexpected and surprising since the electronic spectrum of two graphene bilayers with matched concentrations of electrons and holes is symmetric, as it is clearly seen in Fig.~\ref{Fig2}. The symmetry can be broken by 
Coulomb impurities if most of them are of the same charge. For example positive charges (ionized donors)
provide repulsive scattering for holes and attractive scattering for electrons. 
Our model takes the scattering rates for electrons and holes $\gamma_{\mathrm{t}(\mathrm{b})}$ to be 
momentum and energy independent and ignores these common complications. 
An asymmetry can be introduced can be introduced in phenomenological way by making the assumption that 
Cooper pair-breaking time is energy dependent $\gamma(\omega)$. 
Approximating it by a linear function does predict an asymmetry of the tunneling conductance 
that grows with decreasing temperature, but the shape of the experimental curves is not captured by this simple ansatz.
To clarify whether or not the observed asymmetry can be explained by the presence of charge impurities,
a more microscopic description of their scattering characteristics is needed and this
is outside of the scope of the present work. 
Secondly it is not clear whether or not our theory can capture the dependence of tunneling conductance on 
magnetic field since more experimental data is needed.  Comparison with data obtained at $T\approx 1.5 \;\hbox{K}$ 
suggests that the theory considerably overestimates the effect of magnetic field. Nevertheless, 
at such temperatures the system is in the paired state or in the critical regime that is outside of the applicability range of the theory of Gaussian fluctuations. This discrepancy can be due to other reasons. Bilayer graphene as other two-dimensional systems have long-range density variations and if the corresponding length is smaller than $\hbar/Q_\mathrm{B}$ the effect of the magnetic can not be reduced just to the relative shift of dispersion for electrons from different layers. To better understand capabilities of the theory more experimental data is needed.  

To conclude, we have developed a theory of the fluctuational internal Josephson effect in the system of two closely spaced graphene bilayers. The presence of valley and sublattice degrees of freedom 
provides three competing electron-hole channels for both intra-valley and inter-valley Cooper pairs. 	We show that three channels are nearly independent and have different critical temperatures of the condensation and sublattice structures. The observed enhancement of the tunneling conductance can be explained only by the presence of competing channels that dominate in different temperature ranges. The theory reasonably captures the dependence of the conductance on temperature, voltage bias between bilayers and electron-hole imbalance. We also argue that the enhancement is much stronger for inter-valley tunneling than for the intra-valley one that has been reported recently. We also discuss how to distinguish the uniform state and the FFLO state with finite Cooper pair momentum that can be stabilized in the system by an electron-hole imbalance.

\begin{acknowledgments}
This work was supported by the Army Research Office under Award W911NF-17-1-0312, by ARO MURI 3004628717, and by the Welch Foundation under grant F-1473. DKE acknowledges support from the Australian Research Council Centre of Excellence in Future Low-Energy Electronics Technologies (FLEET). 
\end{acknowledgments}

\bibliography{EHTunnelingBibliography}

\begin{thebibliography}{88}
\bibitem [{Com({\natexlab{a}})}]{CommentCoulombDrag}%
  \BibitemOpen
  \href@noop {} {\enquote {\bibinfo {title} {There are a number of recent
  experiments with double layer electron-hole systems based on conventional
  semiconductors~\cite{ExcitonExp1,ExcitonExp2, ExcitonExp3, ExcitonExp4} and
  novel single-atomic-layer ones~\cite{ExcitonExpGraphene1} that demonstrate an
  anomalous low-temperature behavior in the coulomb drag effect. while there
  are strong evidences that it appears due to interlayer electron-hole
  correlations, a detailed understanding of the anomalous behavior is still
  lacking.}}\ } ({\natexlab{a}})\BibitemShut {NoStop}%
\bibitem [{Com({\natexlab{b}})}]{CommentExcitonPolariton}%
  \BibitemOpen
  \href@noop {} {} ({\natexlab{b}}),\ \bibinfo {note} {it should be noted that
  remarkable progress in observation of the spontaneous coherence of
  exciton-polaritons in semiconductor microcavities has been recently achieved
  (See Refs.~\cite{ExcitonPolariton1, ExcitonPolariton2,ExcitonPolariton3} for
  a review and references therein). Exciton-polariton condensates are close
  relatives of the electron-hole one detected in double bilayer graphene.
  Nevertheless, a maintenance of their coherence requires optical pumping,
  while the latter is fully equilibrium phenomenon.}\BibitemShut {Stop}%
\bibitem [{Com({\natexlab{c}})}]{CommentRs}%
  \BibitemOpen
  \href@noop {} {} ({\natexlab{c}}),\ \bibinfo {note} {this definition of
  $r_\mathrm{s}$ implies the low-energy model with quadratic spectrum that is
  used in our work. This model is applicable if energies of electrons or holes
  ($\epsilon_\mathrm{F}$ and $u$) are far below the interlayer hybridization
  energy $t_\mathrm{BG}\approx 0.4 \; \hbox{eV}$ within graphene bilayer. In
  the experimental setup~\cite{TutucExperiment} Fermi energies and the gap do
  not exceed values $\epsilon_\mathrm{F}\sim 20\;\hbox{meV}$, and $2u\approx
  6.6\, \hbox{meV} $ that is why the condition is well satisfied. It should be
  noted that at much larger $2 |u|\sim 60~\hbox{meV}$ the spectrum becomes
  non-parabolic and has the Mexican-hat shape. This enhances the role of
  interactions and is very important in the strong coupling
  regime~\cite{NeilsonNew}.}\BibitemShut {Stop}%
\bibitem [{Est()}]{Estimations}%
  \BibitemOpen
  \href@noop {} {}\bibinfo {note} {For numerical estimates we use the following
  set of parameters: The dielectric constant $\kappa=7.2$ of $\hbox{WSe}_2$
  that plays the role of the spacer separating graphene bilayers; its thickness
  $d\approx 2\; \hbox{nm}$; effective mass in bilayer graphene $m=0.04 m_0$
  with $m_0$ to be the bare electron mass; thickness of bilayer graphene
  $d_\mathrm{BG}=0.3\;\hbox{nm}$.}\BibitemShut {Stop}%
\bibitem [{Com({\natexlab{d}})}]{CommentSusceptibility}%
  \BibitemOpen
  \href@noop {} {} ({\natexlab{d}}),\ \bibinfo {note} {the Cooper pair
  susceptibility can be defined as $\chi_\mathrm{\Delta}(\tau,\vec{q})=-\langle
  T_\mathrm{M} \Delta(\tau,\vec{q}) \Delta^+(0,\vec{q})\rangle$, where
  $\hat{\Delta}$ is the matrix order parameter of the electron-hole condensate
  in the sub-lattice space. The latter can be found in Appendix A.}\BibitemShut
  {Stop}%
\bibitem [{Com({\natexlab{e}})}]{CommentCooperPropagator}%
  \BibitemOpen
  \href@noop {} {} ({\natexlab{e}}),\ \bibinfo {note} {"The real part of the
  Cooper propagator can be recalculated with the help of Kramers-Kronig
  relations and thus can also been extracted from tunneling
  experiments."}\BibitemShut {NoStop}%
\bibitem [{Com({\natexlab{f}})}]{CommentSaddlePoint}%
  \BibitemOpen
  \href@noop {} {} ({\natexlab{f}}),\ \bibinfo {note} {the saddle point is
  trivial $\langle \hat{\Delta}\rangle=0$ only in the absence of tunneling
  between graphene bilayers. It can be approximated by the trivial if
  $|\hat{t}|^2 \ll \langle |\hat{\Delta}|^2 \rangle$.}\BibitemShut {Stop}%
\end{thebibliography}

\section{Appendixes}

\appendix
\section{A. Bethe-Salpether equation}

Here we present a detailed derivation of Eq.~(\ref{GammaEquation}) from the paper. The Bethe-Saltpeter equation that is illustrated in Fig.~3 can be written as follows 
\begin{widetext}
	\begin{equation}
	\label{BSAp1}
	\Gamma^{\sigma'_\mathrm{t} \sigma_\mathrm{t}}_{\sigma'_\mathrm{b} \sigma_{\mathrm{b}}}(i p_n, \vec{p'}, \vec{p})=U_{\vec{p}'-\vec{p}} \delta^{\sigma'_\mathrm{t} \sigma_\mathrm{t}}_{\sigma'_\mathrm{b} \sigma_{\mathrm{b}}} + T \sum_{i\omega_n \vec{p}''} U_{\vec{p}'-\vec{p}''} G^\mathrm{t}_{\sigma'_t \sigma''_t}(i p_n+i\omega_n,\vec{p}''_+
	) G^\mathrm{b}_{\sigma''_\mathrm{b} \sigma'_\mathrm{b}}(i\omega_n,\vec{p}''_-) 
	\Gamma^{\sigma''_\mathrm{t} \sigma_\mathrm{t}}_{\sigma''_\mathrm{b} \sigma_{\mathrm{b}}}(i p_n, \vec{p}'', \vec{p}).
	\end{equation}
\end{widetext}
Here $\vec{p}_\pm=\vec{p}\pm\vec{q}/2$; $\omega_n=(2n+1)\pi/ T$ and $p_n=2 n\pi / T$ are fermionic and bosonic Matsubara frequencies. The electron Green function $\hat{G}^{\mathrm{t}(\mathrm{b})}(\i \omega_n,\vec{p})$ in the sublattice space can be presented as follows
\begin{widetext}
	\begin{equation}
	G^\mathrm{t}_{\sigma' \sigma}(i \bar{\omega}^\mathrm{t}_n,\vec{p}) = \frac{\langle \sigma'|  \mathrm{t} \mathrm{c} \vec{p}\rangle \langle \mathrm{t} \mathrm{c} \vec{p}|\sigma \rangle}{i \omega_n-\epsilon_{\mathrm{c}\vec{p}}+i\gamma_\mathrm{t}\mathrm{sgn}[\omega_n]}, \quad \quad \quad \quad G^\mathrm{b}_{\sigma' \sigma}(i \omega_n,\vec{p}) = \frac{\langle \sigma'|  \mathrm{b} \mathrm{v} \vec{p}\rangle \langle \mathrm{b} \mathrm{v} \vec{p}|\sigma \rangle}{i \omega_n-\epsilon_{\mathrm{v}\vec{p}}+i \gamma_\mathrm{b}\mathrm{sgn}[\omega_n]}.
	\end{equation}
\end{widetext}
Here $\gamma_{\mathrm{t}(\mathrm{b})}$ are scattering rates for electrons (holes). We have also neglected the presence of the valence band in the layer with excess of electrons and the presence of the conduction band in the layer with the excess of holes. The product of Green functions that appears in (\ref{BSAp1}) can be written with as follows
\begin{widetext}
	\begin{equation*}
	G^\mathrm{t}_{\sigma'_t \sigma''_t}(i p_n+i\omega_n,\vec{p}_+
	) G^\mathrm{b}_{\sigma''_\mathrm{b} \sigma'_\mathrm{b}}(i\omega_n,\vec{p}_-)=M^{\sigma'_t \sigma''_t}_{\sigma'_b \sigma''_b}(\vec{p}'') C(ip_n,\vec{q}, i\omega_n,\vec{p}).
	\end{equation*}	 
\end{widetext}
Here the matrix form-factor 
\begin{equation*}
M^{\sigma'_t \sigma''_t}_{\sigma'_b \sigma''_b}(\vec{p})=\langle \sigma_\mathrm{t}'|  \mathrm{t} \mathrm{c} \vec{p}_+\rangle \langle \mathrm{t} \mathrm{c} \vec{p}_+|\sigma_\mathrm{t}'' \rangle \langle \sigma''_\mathrm{b}|  \mathrm{t} \mathrm{v} \vec{p}_-\rangle \langle \mathrm{b} \mathrm{v} \vec{p}_-|\sigma'_\mathrm{b} \rangle.
\end{equation*}
reflects the chiral nature of charge carriers in bilayer graphene, while $C(ip_n,\vec{q},i\omega_n, \vec{p})$ contains information only about energy spectrum for electrons and holes and is given by
\begin{widetext}
	\begin{equation}
	C(ip_n,\vec{q},i\omega_n,\vec{p})=\frac{1}{(i \omega_n+i p_n-\epsilon_{\mathrm{c}\vec{p}_+} + i \gamma_\mathrm{t}\mathrm{sgn}[\omega_n+p_n])(i \omega_n-\epsilon_{\mathrm{v}\vec{p}_-}+i \gamma_\mathrm{b} \mathrm{sgn}[\omega_n])}
	\end{equation}
\end{widetext}
In the weak coupling regime that we consider in the paper pairing correlations do appear in the vicinity of Fermi lines for electrons and holes. As a result the vertex $\hat{\Gamma}(i p_n, \phi_{\vec{p}'}, \phi_{\vec{p}})$, the Fourier transform of interactions $U(\phi_{\vec{p}'}-\phi_\vec{p})$, and the form-factor $\hat{M}(\phi_\vec{p})$ can be safely approximated by their values at the Fermi level ($|\vec{p}|=p_\mathrm{F}$ and $|\vec{p}'|=p_\mathrm{F}$) and depend only on the corresponding polar angles ($\phi_\vec{p}$ and $\phi_{\vec{p}'}$). After decomposition over multipole momenta  $\hat{\Gamma}_{l',l}$, $U_l$ and $\hat{M}_l$ the equation (\ref{BSAp1}) becomes algebraic and can be presented in a compact form
\begin{equation}
\label{BSAp2}
\hat{\Gamma}_{l'l}=U_{l'} \delta_{l'l} \hat{1}+ \sum_{l_1 l_2} U_{l'} M_{l'-l_1-l_2} \Pi_{l_2}(ip_n, \vec{q}) \hat{\Gamma}_{l_2,l}.  
\end{equation}
Here $\Pi_l(i p_n, \vec{q})$ is given by
\begin{equation*}
\Pi_l(i p_n, \vec{q})=T\sum_{i \omega_n, \vec{p}}e^{-i l \phi_\vec{p}}C(ip_n,\vec{q},i\omega_n, \vec{p})
\end{equation*}
Its $s$-wave component $\Pi_0(i p_n, \vec{q})\equiv\Pi(ip_n,\vec{q})$ has the ultraviolet logarithmic divergence that is usually present in the weak coupling pairing theories and coincides with the single-step pair propagator in the Cooper ladder sum of a bilayer system without sublattice degrees of freedom. Its detailed derivation can be found in textbooks~\cite{Varlamov,LarkinVarlamov} and its explicit expression is presented in the main part as Eq.~(\ref{Pi}). Therefor all information about the chiral nature of the bilayer graphene charge carriers is hidden in the nontrivial matrix form-factor $\hat{M}_l$. At finite $l$ the value $\Pi_l(i p_n, \vec{q})$ is nonzero only at finite Cooper pair momentum and are much smaller than $\Pi_0(i p_n, \vec{q})$ and can be neglected. As a result, the Eq.~(\ref{BSAp2}) reduces to Eq.~(\ref{GammaEquation}) from the main text of the paper.  

\section{B. Inter-valley alignment and identification of the FFLO state}
\begin{figure}[b]
	\begin{center}
		\includegraphics[width=1.\columnwidth]{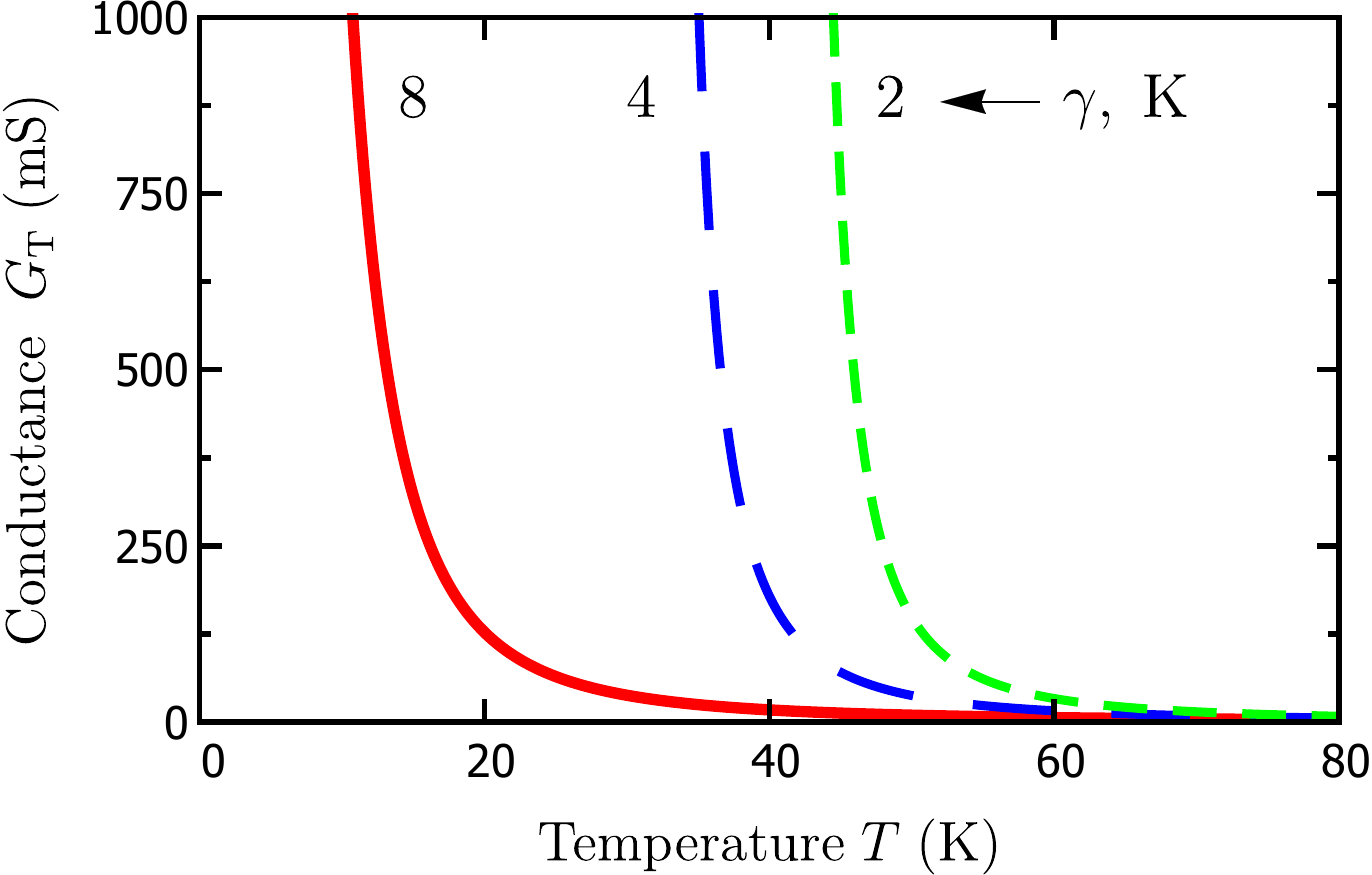}
		\caption{The temperature dependence of the inter-valley tunneling conductance $G_\mathrm{T}$ between graphene bilayers at zero voltage bias ($V=0$). Three curves correspond to pair breaking rates $\gamma=2,\, 4,\,\hbox{and}\; 8 \;\hbox{meV}$. The pair breaking rate induced by scattering at impurities reduces the critical temperature $T_0$ of electron-hole condensation, but weakly effect the critical behavior above $T_0$.}
		\label{FigApA1}
	\end{center}\vspace{-0.5cm}
\end{figure} 

\begin{figure}[t]
	\begin{center}
		\includegraphics[trim=0cm 6cm 0.4cm 0cm, clip, width=0.9\columnwidth]{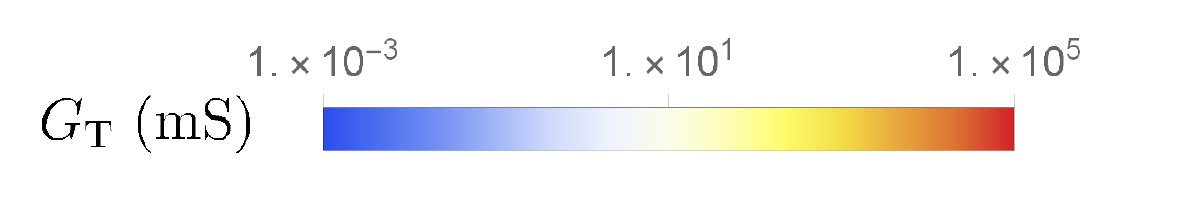}
		\includegraphics[trim=0.1cm 0cm 0.1cm 0.1cm, clip, width=0.9\columnwidth]{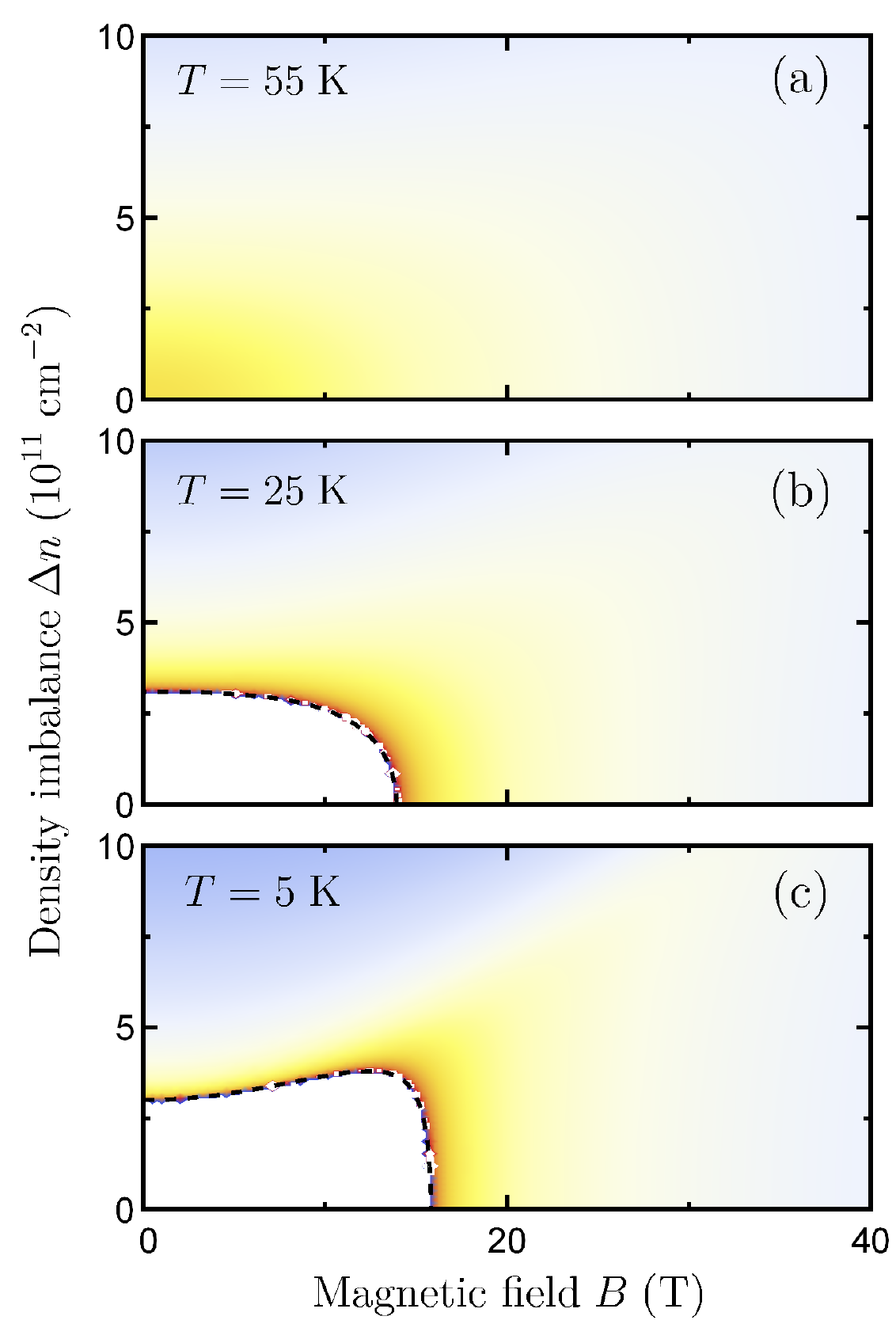}
		\caption{The dependence of tunneling conductance $G_\mathrm{T}$ on in-plane magnetic $B$ and the electron-hole density imbalance $\Delta n$. Three subplots correspond to $T=45 \; \hbox{K}$ $\hbox{(a)}$, $25 \; \hbox{K}$ $\hbox{(b)}$ and $5 \; \hbox{K}$ $\hbox{(c)}$. The dashed line corresponds to the phase boundary of equilibrium electron-hole paired state. In ($\hbox{b}$) and ($\hbox{c}$) the conductance achieves maximum at finite value of magnetic field $B$ that demonstrates that fluctuating Cooper pairs with finite momentum $\vec{Q}\approx \vec{Q}_\mathrm{B}$ are the most intensive and the system is in the vicinity of the instability to the FFLO state. The presence of a kink in the phase boundary in ($\hbox{c}$) clearly demonstrates that the system is unstable towards the equilibrium FFLO state.}
		\label{FigApA2}
	\end{center}
\end{figure}
In the main text of the paper calculations for intra-valley tunneling ($\theta= 0$) are discussed, while ones for the inter-valley alignment ($\theta= \pi/3$) are presented here. The tunneling conductance is dominated by the channel $11\hbox{-}22$ because of fluctuating Cooper pairs in the dominating channel involve correlations at adjacent sublayers. The tunneling conductance can be approximated as Eq.~(\ref{GTInter}). We use the same set of parameters that has been used above to fit the experimental data except for pair-breaking rate $\gamma$. For the latter we use $\gamma=2,\, 4,\,\hbox{and}\; 8 \;\hbox{meV}$. The first two values correspond to cleaner samples compared to ones that have been studied experimentally \cite{TutucExperiment}. The temperature dependence of tunneling conductance is presented in Fig.~\ref{FigApA1}. Its enhancement of the conductance is considerably stronger than that for the inter-valley tunneling. The pair-breaking rate $\gamma$ determines the critical temperature of pair condensation, but weakly influences the temperature dependence of tunneling conductance above it.

For the pair-breaking rate $\gamma=2\;\hbox{meV}$ the critical temperature is $T_0\approx 42 \; \hbox{K}$. According to the phase diagram Fig.~\ref{Fig4} the ratio $\gamma/T_0\approx 0.52$ is small enough to stabilize the Fulde-Ferrell-Larkin-Ovchinnikov (FFLO) state by the electron-hole imbalance. In-plane magnetic filed results in a relative shift of electronic dispersions between layers and makes it possible to probe fluctuating Cooper pairs with finite momentum $Q_\mathrm{B}=ed B_{||}/\hbar c$. The dependence of tunneling conductance at zero voltage bias on the electron-hole imbalance and magnetic field at temperature $T=55\; \hbox{K}$ is presented in Fig.~\ref{FigApA2}-a. The tunneling conductance has a monotonic dependence demonstrating that the system is far away from the instability to the FFLO state and Cooper pair fluctuations with zero momentum $\vec{Q}=0$ are the most intensive. The dependence at $T=25\; \hbox{K}$ is presented in Fig.~\ref{FigApA2}-b with a dashed line that denotes a phase boundary of a paired state. Since the critical imbalance required to suppress the pairing instability decreases with magnetic field the paired state is the uniform BCS one. Nevertheless, the tunneling conductance achieves maximum at finite value of magnetic field that shows that the systems is close to the FFLO state and fluctuations with finite Cooper momentum are the most intensive. The dependence at $T=5\; \hbox{K}$ is presented in Fig.~\ref{FigApA2}-c. The dashed line that denotes the phase boundary is non-monotonous and has a kink at finite value of magnetic field. It clearly demonstrates that the equilibrium FFLO state is stabilized by the electron-hole imbalance.        

\section{C. Bosonic picture of the fluctuational internal Josephson effect}
Here we present a bosonic picture of the fluctuational internal Josephson effect that can be employed with a help of the field integral formalism. We consider the contact interactions $U$ between electrons and holes that correspond to the truncation of all multipole momenta of interactions except the $s$-wave one. In the main text of the paper we have demonstrated that they are unimportant. The  correlation function of tunneling operators $\chi(\omega,\vec{q})$ that defines the tunneling conductance $G_\mathrm{T}$ according to Eq.~(\ref{GT1}) can be extracted from the corresponding imaginary time correlation function $\chi(\tau,\vec{r})$ given by     
\begin{equation}
\label{TunnelingResponse1}
\chi(\tau,\vec{r})=\frac{1}{Z} \frac{\delta^2 Z}{\delta \Lambda_{\tau,\vec{r}} \delta\Lambda^+_{0,0}}.
\end{equation}
Here $Z[\Lambda^+_{\tau,\vec{r}},\Lambda_{\tau,\vec{r}}]$ is the statistical sum with an auxiliary bosonic field $\Lambda_{\tau,\vec{r}}$ introduced to the action $S$ of the system as follows 
\begin{widetext}
	\begin{equation*}
	\label{S1ap}
	S=\int_0^\beta d\tau \int \vec{dr} \left[ 
	\hat{\psi}_{\mathrm{t}}^+ (\partial_\tau+\hat{h}_{\mathrm{t} \vec{r}} - \mu_\mathrm{t}) \hat{\psi}_{\mathrm{t}} + \hat{\psi}_{\mathrm{b}}^+ (\partial_\tau+\hat{h}_{\mathrm{b} \vec{r}} + \mu_\mathrm{b} ) \hat{\psi}_{\mathrm{b}} + \hat{\psi}_{\mathrm{b}}^+ \,\Lambda^+\hat{t}^+\, \hat{\psi}_{\mathrm{t}} + \hat{\psi}_{\mathrm{t}}^+ \hat{t} \Lambda\, \hat{\psi}_{\mathrm{b}} +  U\psi^+_{\mathrm{t} s_\mathrm{t}} \psi^+_{\mathrm{b} s_\mathrm{b}} \psi_{\mathrm{b} s_\mathrm{b}} \psi_{\mathrm{t} s_\mathrm{t}}\right].
	\end{equation*}
\end{widetext}	
Here $\hat{\psi}_{\mathrm{t}}\equiv \hat{\psi}_{\mathrm{t}\tau \vec{r}}$ and $\hat{\psi}_{\mathrm{b}}\equiv \hat{\psi}_{\mathrm{b}\tau \vec{r}}$ are spinor fermionic fields for electrons from top (t) and bottom (b) layers with labeling described in Sec.~II\ref{SecII}. If they are integrated out and the corresponding action is expanded in the lowest order in tunneling matrix elements $\hat{t}$, the Fourier transform $\chi_q$ do appears in the action as follows $S=-\sum_{q} \chi_q |\Lambda_q|^2$. Here  $q=\left\{q_n,\vec{q}\right\}$ with bosonic Matsubara frequency $q_n=2\pi T n$. For noninteracting electrons and holes calculations are straightforward and result in  
\begin{equation}
\label{Chi0ap}
\chi^0_q =\hat{t}^+ \hat{M}_0 N_\mathrm{F}\Pi_q \, \hat{t}. 
\end{equation}
After analytical continuation we get the Eq.~(\ref{Chi0}) from the main text. In the case of interacting electrons and holes it is instructive start with the Hubbard-Stratonovich transformation. It eliminates interactions but introduces the bosonic field $\hat{\Delta}\equiv\hat{\Delta}_{\tau \vec{r}}$ corresponding to electron-hole Cooper pairs. The action $S$ is modified as follows
\begin{widetext}
	\begin{equation}
	\label{S2ap}
	S=\int_0^\beta d\tau \int \vec{dr} \left[ 
	\hat{\psi}_{\mathrm{t}}^+ (\partial_\tau+\hat{h}_{\mathrm{t} \vec{r}} - \mu_\mathrm{t}) \hat{\psi}_{\mathrm{t}} + \hat{\psi}_{\mathrm{b}}^+ (\partial_\tau+\hat{h}_{\mathrm{b} \vec{r}} + \mu_\mathrm{b} ) \hat{\psi}_{\mathrm{b}} + \hat{\psi}_{\mathrm{b}}^+ \,\hat{\Delta}'^+ \, \hat{\psi}_{\mathrm{t}} + \hat{\psi}_{\mathrm{t}}^+ \hat{\Delta}'\, \hat{\psi}_{\mathrm{b}} + \frac{1}{U} \mathrm{tr}\left[\hat{\Delta}^+ \hat{\Delta}\right]\right].
	\end{equation}
\end{widetext}	
where $\hat{\Delta}'_{\tau \vec{r}}=\hat{\Delta}_{\tau \vec{r}}+\hat{t}\, \Lambda_{\tau \vec{r}}$. Above the critical temperature $T_0$ of the electron-hole pairing the saddle point of the action is trivial $\langle \hat{\Delta}\rangle=0$ and the field $\hat{\Delta}$ corresponds to Cooper pair fluctuations~\cite{CommentSaddlePoint}. In the wide temperature range $\Delta T _{Gi}\lesssim\Delta T\sim T_0$  outside the critical regime $\Delta T \lesssim \Delta T _{Gi}$ fluctuations can be approximated by the noniteracting Gaussian theory. Here $\Delta T_{Gi}=Gi\; T_0$ with Ginzburg number $Gi=T_0/E_\mathrm{F}$ calculated in Appendix D. Integrating out fermions and expanding the action up to the second order in the bosonic field $\hat{\Delta}'$ results in 
\begin{widetext}
	\begin{equation}
	\label{S3ap}
	S=\sum_q\left[\frac{\hat{\Delta}_q^+ \hat{\Delta}_q}{U}  -\hat{\Delta}'^+_q  \hat{M}_0 \Pi_q \hat{\Delta}_q'\right]= \sum_q\left[\hat{\Delta}_q^+ \hat{\Gamma}_q^{-1} \hat{\Delta}_q -\hat{\Delta}_q^+ \hat{M}_0 \Pi_q t \Lambda_q -  \Lambda^+_q \hat{t}^+ \hat{M}_0 \Pi_q \hat{\Delta}_q-\Lambda^+_q \chi^0_q \Lambda_q   \right].
	\end{equation}
\end{widetext}	

Here all matrices are in the compact representation, and $\hat{\Gamma}_q$ is the scattering vertex calculated in the Sec. III of the paper. Some of its components vanish at the critical temperature $T_0$ of the electron-hole Cooper pairing. The last term corresponds to the response $\chi^0_q$ function for noninteracting electrons and holes that is given by Eq.(~\ref{Chi0ap}). The action represents the bosonic picture of the Josephson effect and is valid outside the weak coupling regime. The action~(\ref{S3ap}) is quadratic in the bosonic field $\Delta_{q}$ and after its integration we get the tunneling response function $\chi_q$ to be given by
\begin{equation}
\label{Chiap}
\chi_q=\chi^0_q+ \hat{t}^+\hat{M}_0 \Pi_q \hat{\Gamma}^{-1}_q \hat{M}_0 \Pi_q \hat{t}   
\end{equation}
With a help of Eqs.~(\ref{GammaKK}) and~(\ref{MKK}) we get the tunneling response function for the intra-valley tunneling~(\ref{ChiKK}). In the same way with a help of Eqs.~(\ref{GammaKKp}) and~(\ref{MKKp}) we get the response function for the inter-valley one~(\ref{ChiKKp}). 
\section{D. Ginzburg criterion}
The developed theory of the fluctuational internal Josephson effect implies that Cooper pair fluctuations are Gaussian and do not interact with each other. The interactions can be safely omitted in the wide range of temperatures $\Delta T\sim T_0$ except the critical region $\Delta T \lesssim \Delta T_\mathrm{Gi}$ where fluctuations are overgrown are strongly interfere with each other. The range $\Delta T_\mathrm{Gi}$ can be estimated from the Ginzburg criterion~\cite{Ginzburg} that compares the contribution of Gaussian fluctuations to the heat capacity $C_\mathrm{FL}= T_0/4\pi \xi^2 \Delta T$ with predictions of the mean-field theory below the critical temperature $C_\mathrm{MF}=N_\mathrm{F}/g T_0=\epsilon_\mathrm{F}/4\pi \xi^2 T_0$. Here $g=4 \xi^2/(\hbar v_\mathrm{F})^2$ is the strength of contact interactions between fluctuating Cooper pairs neglected so far. The contribution of fluctuations $C_\mathrm{MF}$ grows with decreasing of temperatures and dominates in the temperature range $\Delta T_{Gi}=Gi T_0$ with Ginzburg number $Gi=T_0/E_\mathrm{F}$. It does not depend explicitly on the pair-breaking rate $\gamma$ but only on the critical temperature $T_0$.  It should be noted that the Ginzburg criterion can be derived microscopically in a more strict way by explicit analysis of the role of interactions between fluctuating Cooper pairs ~\cite{Dodgson}.

\end{document}